\documentclass[twocolumn,aps,prl,amsmath,showpacs,amssymb,floatfix]{revtex4}
\usepackage{tabularx}
\usepackage{bm}
\usepackage{subfigure}
\usepackage{euscript}
\usepackage{graphicx}
\usepackage{color}
\usepackage[colorlinks=true,linkcolor=blue]{hyperref}%
\usepackage{exscale}
\usepackage{amsbsy}
\usepackage{docs}
\usepackage{amsfonts,amsmath,amssymb}
\usepackage{dcolumn}

\begin{document}

\title{Study on the tunneling spectroscopy of $N-pS$ junction and $N-hS$ junction}

\author{Zhongbo Yan}
\author{Shaolong Wan}
\email[]{slwan@ustc.edu.cn}
\affiliation{Institute for Theoretical
Physics and Department of Modern Physics University of Science and
Technology of China, Hefei, 230026, P. R. China}
\date{\today}

\begin{abstract}
We study the complete tunneling spectroscopy of a normal
metal/$p$-wave superconductor junction ($N-pS$) and normal
metal/heterostructure superconductor junction ($N-hS$) by
Blonder-Tinkham-Klapwijk (BTK) method. We find that, for $p$-wave
superconductor with non-trivial topology, there exists a quantized
zero-bias conductance peak stably, for heterostructure
superconductor with non-trivial topology, the emerging zero-bias
conductance peak is non-quantized and usually has a considerable
gap to the quantized value. Furthermore, it is sensitive to
parameters, especially to spin-orbit coupling and the $s$-wave
pairing potential. Results obtained suggest that the observation
of a small zero-bias conductance peak, instead of a quantized
zero-bias conductance peak, in current tunneling experiments can
be a natural result if the spin-orbit coupling turns out to be
several times smaller than the reported one. Results obtained also
suggest that both a stronger spin-orbit coupling and proximity
$s$-wave superconductor with relative weaker pairing potential can
produce a much more striking zero-bias conductance peak (compared
to the experiments), even an almost quantized one. As $s$-wave
superconductors are common in nature, the prediction can be
verified within current experiment ability.
\end{abstract}


\maketitle

\section{Introduction}
\label{sec1}

Because of hosting exotic non-abelian zero modes \cite{D. A.
Ivanov} which have great potential in topological quantum
computation \cite{A. Kitaev1, S. Das. Sarma, C. Nayak}, $p$-wave
superconductor either in two dimension \cite{N. Read} or in one
dimension \cite{A. Kitaev2} has raised strong and lasting
interests for more than a decade. Although there is no definite
confirmation of $p$-wave superconductor in solid state physics
\cite{A. P. Mackenzie, C. Kallian}, several groups \cite{Liang
Fu1, Jay D. Sau, Roman M. Lutchyn, Y. Oreg, J. Alicea1, A. C.
Potter}, based on proximity effect, have proposed a series of
heterostructures whose common elements are spin-orbit coupling,
$s$-wave superconducting order and Zeeman field and found, by
tuning parameters, the upper band of the system can be projected
away, and the Copper pairs formed in the lower band is ``effective
$p$-wave", and for such an  ``effective $p$-wave" superconductor,
the zero modes known as Majorana bound states emerge at defects or
the boundary of the system. Such heterostructures with non-trivial
topology is known as topological superconductors.

To detect the Majorana bound states, there are mainly three
classes of measurements schemes \cite{J. Alicea2, T. D. Stanescu},
based on tunneling \cite{K. Sengupta, C. J. Bolech, Johan Nilsson,
Liang Fu2, A. Akhmerov, K. T. Law, Karsten Flensberg, Stefan
Walter, Dong E. Liu, A. R. Akhmerov, R. Zitko, Y. Cao, S. Mi, Z.
Wang}, fractional Josephson effects \cite{P. A. Ioselevich, Liang
Jiang, Pablo San-Jose, L. Jiang, Fernando Dominguez} and
interferometry \cite{Colin Benjamin, Chao-Xing Liu, Andrej
Mesaros}. Recently, several tunneling experiments \cite{V. Mourik,
A. Das, M. T. Deng, A. D. K. Finck} based on the proposals of
one-dimensional topological superconductors \cite{Roman M.
Lutchyn, Y. Oreg} were carried out and all these experiments found
that a zero-bias conductance peak, which is taken as a signature
of Majorana bound states, emerges in the tunneling spectroscopy
when the magnetic field along the nanowire exceeds the critical
value, $i.e.$, $B>B_{c}=\sqrt{\Delta^{2}+\mu^{2}}$. However, it is
also noticed that the peak height is quite small, having a big
discrepancy to the theoretical prediction: a quantized zero-bias
peak of height $2e^{2}/h$ (at zero temperature) \cite{K. Sengupta,
K. T. Law, Karsten Flensberg}. The big discrepancy has raised
debate on the origin of the zero-bias conductance peak. To
understand the experiments and clarify the origin of the peak,  a
series of work \cite{E. Prada, Tudor D. Stanescu, C.-H. Lin, E. J.
H. Lee, G. Kells, J. Liu, D. Bagrets, D. I. Pikulin, D. Rainis, D.
Roy} have been carried out, and some of the work  point out that
the non-quantized zero-bias conductance peak can have several
other origins, like Kondo effect \cite{E. J. H. Lee}, smooth end
confinement \cite{G. Kells}, strong disorder \cite{J. Liu, D.
Bagrets, D. I. Pikulin}, and suppression of the superconducting
pair potential at the end of the heterostructure \cite{D. Roy}.
Therefore, a definite confirmation of Majorana bound states is
still lacking. As the heterostructures are proposed to be an
``effective $p$-wave" superconductor, a comparative study of the
tunneling spectroscopy of a normal metal/$p$-wave superconductor
($N-pS$) junction and normal metal/heterostructure superconductor
($N-hS$) (here we name the heterostructure as heterostructure
superconductor, instead of topological superconductor, since it
can also be topologically trivial) junction to see what extent the
``effective" can reach is important and valuable, for both
understanding of the realized experiments and giving some guide
for future experiments.

In this article, according to the BTK method \cite{G. E. Blonder},
we systematically consider the effects due to (i) the length ($L$)
of the system, (ii) interface scattering potential ($H$), (iii)
chemical potential mismatch ($\delta \mu$), to the tunneling
spectroscopy of both junctions, and only consider the effects due
to (iv) spin-orbit coupling ($\alpha$), (v) magnetic field ($B$)
along the wire, (vi) $s$-wave pairing potential, to the tunneling
spectroscopy of $N-hS$ junction. For $N-pS$ junction, we obtain a
complete tunneling spectroscopy analytically, including
probability of normal reflection (an electron reflected as an
electron with the same spin), probability of equal-spin Andreev
reflection (an electron reflected as a hole with the same spin),
and the differential tunneling conductance and their dependence on
system parameters. For $N-hS$ junction, we obtain the complete
tunneling spectroscopy numerically, including probability of
normal reflection, probability of spin-reversed normal reflection
(an electron reflected as an electron with the opposite spin),
probability of spin-reversed Andreev reflection (an electron
reflected as a hole with the opposite spin), probability of
equal-spin Andreev reflection and the differential tunneling
conductance and their dependence on parameters. Although only the
differential tunneling conductance is observable in experiments,
the other reflection coefficients are also very important for us
to understand the underlying tunneling process. Among these
reflection coefficients, spin-reversed Andreev reflection and
equal-spin Andreev reflection are worthy of attention, they can
tell us whether the system favors $s$-wave pairing or $p$-wave
pairing, and give a quantitative estimate of the extent the
``effective" has reached.

The main results obtained, for $N-pS$ junction: (a) When $\mu > 0$
and $L$ is sufficiently long, the zero-bias peak is always
quantized, of height $2e^{2}/h$ ($T=0$) and  independent of $H$,
however, once $\mu_{s}$ crosses zero and becomes negative, $i.e.$,
$\mu_{s}<0$, the zero-bias conductance peak changes into a
conductance dip, with value very close to zero, which indicates
there exists a topological quantum transition when $\mu_{s}$
crosses zero. (b) When $L$ is short, the conductance is very close
to zero and no conductance peak appears both for $\mu_{s} > 0$ and
$\mu_{s} < 0$. However, when $L$ is increased to intermediate
value, a non-quantized conductance peak appears at finite energy
for $\mu_{s} > 0$ and with the increase of $L$, the peak moves
toward to zero-bias voltage with height increasing to the
quantized value. For $N-hS$ junction: (a) With infinity length, we
find when the magnetic field $B$ crosses the critical value
$B_{c}$, a zero-bias peak appears, however, unlike the $p$-wave
case, the peak height is non-quantized, and parameters dependent.
(b) Decreasing the spin-orbit coupling will reduce the height of
the zero-bias conductance peak, and when the spin-orbit coupling
decreases to zero and other parameters keep unchanged, the
zero-bias conductance peak disappears, which indicate the
breakdown of the usual topological criterion for a heterostructure
superconductor. When $L$ is intermediate, similar to $N-pS$
junction, there is also a non-quantized conductance peak appearing
at finite energy for $B > B_{c}$. (c) Adopting the experimental
parameters, we study the effects of the $s$-wave pairing potential
and find that a more striking zero-bias conductance peak favors a
weaker pairing potential. Furthermore, we find that the spin-orbit
coupling several times smaller than the reported one in the
experiment \cite{V. Mourik} which can be taken as a possible
explanation of the quite small zero-bias conductance observed in
the experiment.

The paper is organized as follows: In Sec.\ref{sec2}, we give the
theoretical models for $N-pS$ junction and $N-hS$ junction
explicitly, and based on the BTK method \cite{G. E. Blonder}, we
obtain the tunneling spectroscopy of $N-pS$ junction and $N-hS$
junction under different parameter conditions. In Sec.\ref{sec3},
we give a discussion of the tunneling spectroscopy of $N-pS$
junction and $N-hS$ junction obtained in Sec.\ref{sec2}. We also
conclude the paper at the end of Sec.\ref{sec3}.

\section{Theoretical model}
\label{sec2}

\subsection{$N-pS$ junction}
\label{subsec2a}

\begin{figure}
\subfigure{\includegraphics[width=8.0cm, height=3cm]{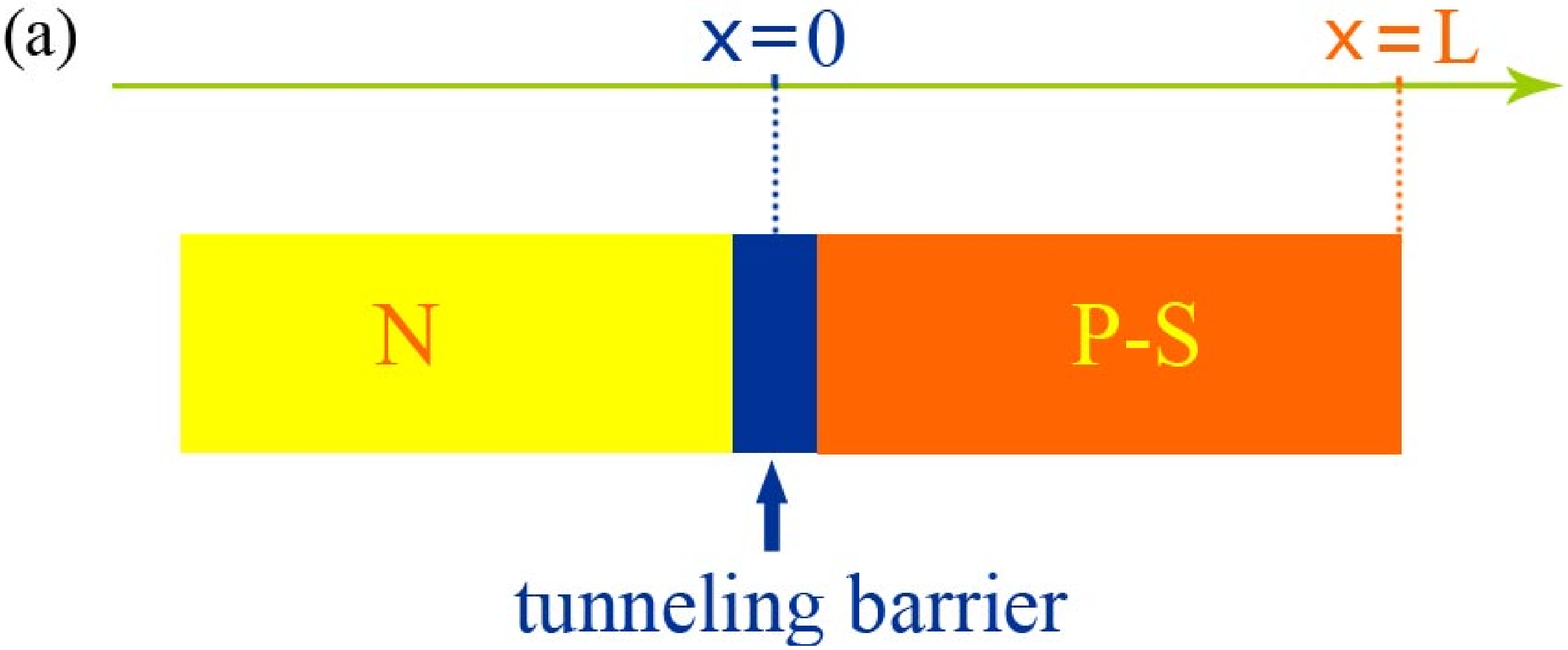}}
\subfigure{\includegraphics[width=8.0cm, height=3.8cm]{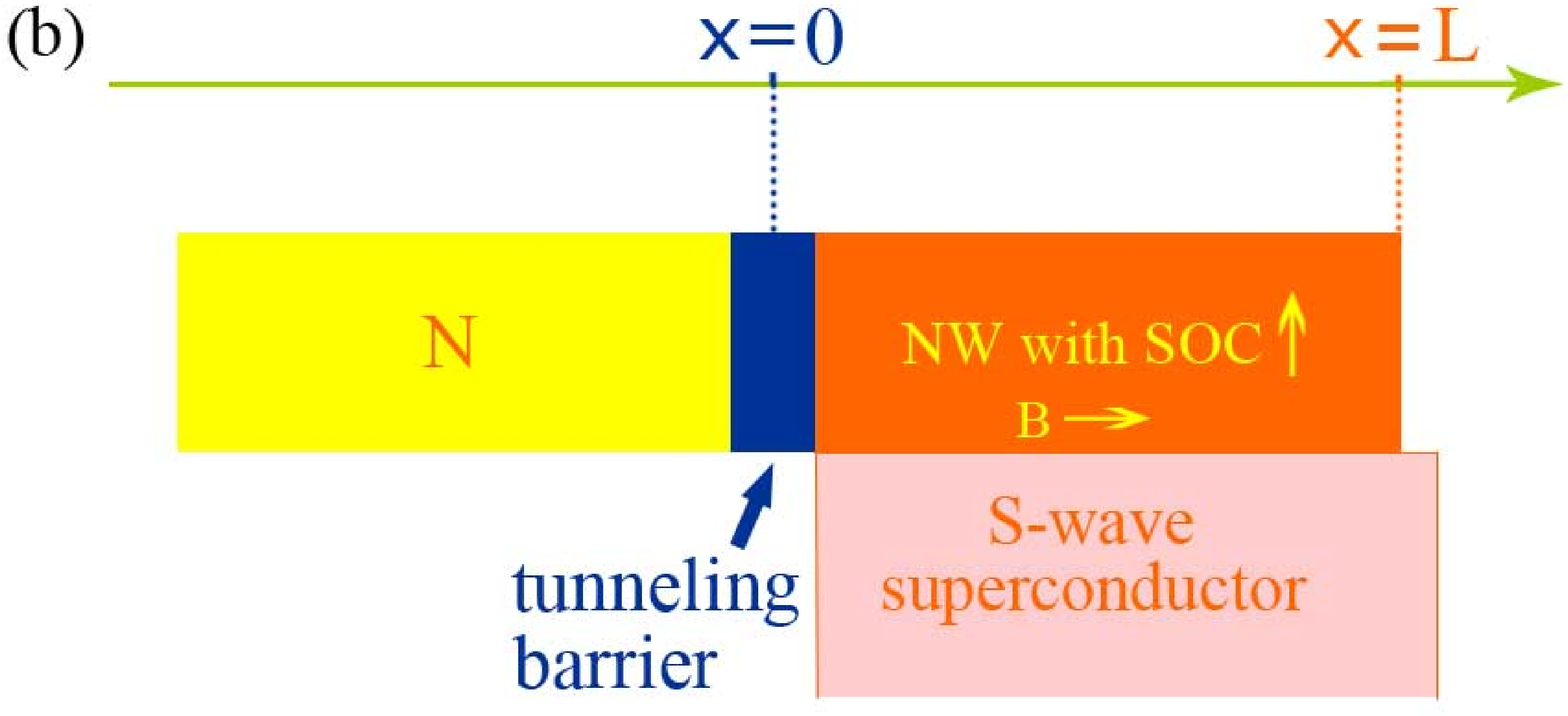}}
\caption{ (color online)(a) normal metal (N)/$p$-wave
superconductor (pS) junction. (b) normal metal (N)/heterostructor
superconductor (hS) junction. $x < 0$, the normal metal, $0 < x <
L$, $p$-wave superconductor or the nanowire (NW). The tunneling
barrier is modeled as $H \delta(x)$, here we broaden its width for
illustration.}
\label{fig1}
\end{figure}

We first consider the one-dimensional $N-pS$ junction shown in
Fig.\ref{fig1}(a). Under the representation
$\Psi^{\dag}(x)=(\psi^{\dag}(x),\psi(x))$, the Hamiltonian is
given as
\begin{eqnarray}
H=\left[-\frac{\hbar^{2}}{2m} \frac{d^{2}}{dx^{2}} - \mu(x) + V(x)
+ H \delta(x) \right] \sigma_{z} + \Delta(x) \sigma_{x}, \label{1}
\end{eqnarray}
where $\vec{\sigma}=(\sigma_{x}, \sigma_{y}, \sigma_{z})$ are
pauli matrices, $H \delta(x)$ is the scattering potential at the
interface, $V(x)$ is potential induced by disorder, external
field, $etc$, for $N-pS$ junction, we set $V(x) = 0$. $\mu(x)$ is
the chemical potential, we set $\mu(x)=\mu_{n}$ for $x<0$ and
$\mu(x) = \mu_{s}$ for $0 < x < L$, $\delta \mu = \mu_{n} -
\mu_{s}$ is the chemical mismatch. $\Delta(x) = -i \Delta
\theta(x) \theta(L-x) \hbar \partial_{x}$ is the pairing
potential, we assume it only appears at $0 < x < L$.

For $0 < x < L$, the $p$-wave superconductor, the Hamiltonian in
momentum space is given as (in the following, we set $\hbar = m =
1$)
\begin{eqnarray}
H_{S} = \left[\frac{k^{2}}{2} - \mu_{s} \right] \sigma_{z} +
\Delta k \sigma_{x}, \label{2}
\end{eqnarray}
which is the continuum form of the Kitaev model \cite{A. Kitaev2}.
Its energy spectrum is $E = \pm \sqrt{(\frac{k^{2}}{2} -
\mu_{s})^{2} + (\Delta k)^{2}}$. As we know, the energy gap is
only closed at $\mu=0$, and for $\mu_{s} > 0$ $(\mu_{s} < 0)$, the
system is topologically non-trivial (trivial).

To study the behavior of tunneling spectroscopy when $\mu_{s}$
goes across the critical point $\mu_{c} = 0$, we consider
$\Delta^{2} > 2 |\mu_{s}|$, where $\mu_{s}$ is very close to
$\mu_{c}$ and the minimum of the energy spectrum is located at $k
= 0$. For a given energy $E$ (here we only consider in-gap states,
$i.e.$, $E < E_{g}/2 = |\mu_{s}|$, as the differential tunneling
conductance at zero-bias voltage is what we are the most
interested in), the wave function in $p$-wave superconductor is
given as
\begin{eqnarray}
\psi_{S}(x)&= c(E)\left( \begin{array}{c}
        iu_{+} \\
       v_{+}
      \end{array} \right) e^{-k_{+}x} + \tilde{c}(E) \left( \begin{array}{c}
        -iu_{+} \\
      v_{+}
      \end{array} \right) e^{k_{+}x} \nonumber \\
&+ d(E)\left( \begin{array}{c}
        iu_{-} \\
       v_{-}
      \end{array} \right) e^{-k_{-}x} + \tilde{d}(E) \left( \begin{array}{c}
        -iu_{-} \\
       v_{-}
      \end{array} \right) e^{k_{-}x}, \label{3}
\end{eqnarray}
where $k_{\pm} = \sqrt{2 (\Delta^{2} - \mu_{s}) \pm 2 \sqrt{E^{2}
+ \Delta^{2}(\Delta^{2} - 2\mu_{s})}}$, $u_{\pm} = \Delta
k_{\pm}$, and $v_{\pm} = E + \frac{k_{\pm}^{2}}{2} + \mu$. $c(E)$,
$\tilde{c}(E)$, $d(E)$ and $\tilde{d}(E)$ are energy-dependent
coefficients, and when $L = \infty$, $\tilde{c}(E) = \tilde{d}(E)
= 0$, it is obvious that the wave function is localized at the
left end of the wire.

For $x < 0$, the normal metal lead, the Hamiltonian is,
\begin{eqnarray}
H_{N}=\left[\frac{k^{2}}{2} - \mu_{n} \right] \sigma_{z},
\label{4}
\end{eqnarray}
here we keep $\sigma_{z}$ for convenience. We consider that an
electron is injected from the normal lead into the $p$-wave
superconductor, and the wave function in the normal lead is given
as
\begin{eqnarray}
\psi_{N}(x) = \left( \begin{array}{c}
        e^{2iq_{e}x} + b(E) \\
      0
      \end{array} \right) e^{-iq_{e}x} + a(E) \left( \begin{array}{c}
        0 \\
      1
      \end{array} \right) e^{iq_{h}x}, \label{5}
\end{eqnarray}
where $q_{e,h} = \sqrt{2(\mu_{n} \pm E)}$. $a(E)$ and $b(E)$
denote equal-spin Andreev reflection amplitude and normal
reflection amplitude, respectively.

\begin{figure}
\subfigure{\includegraphics[width=4.0cm, height=3.0cm]{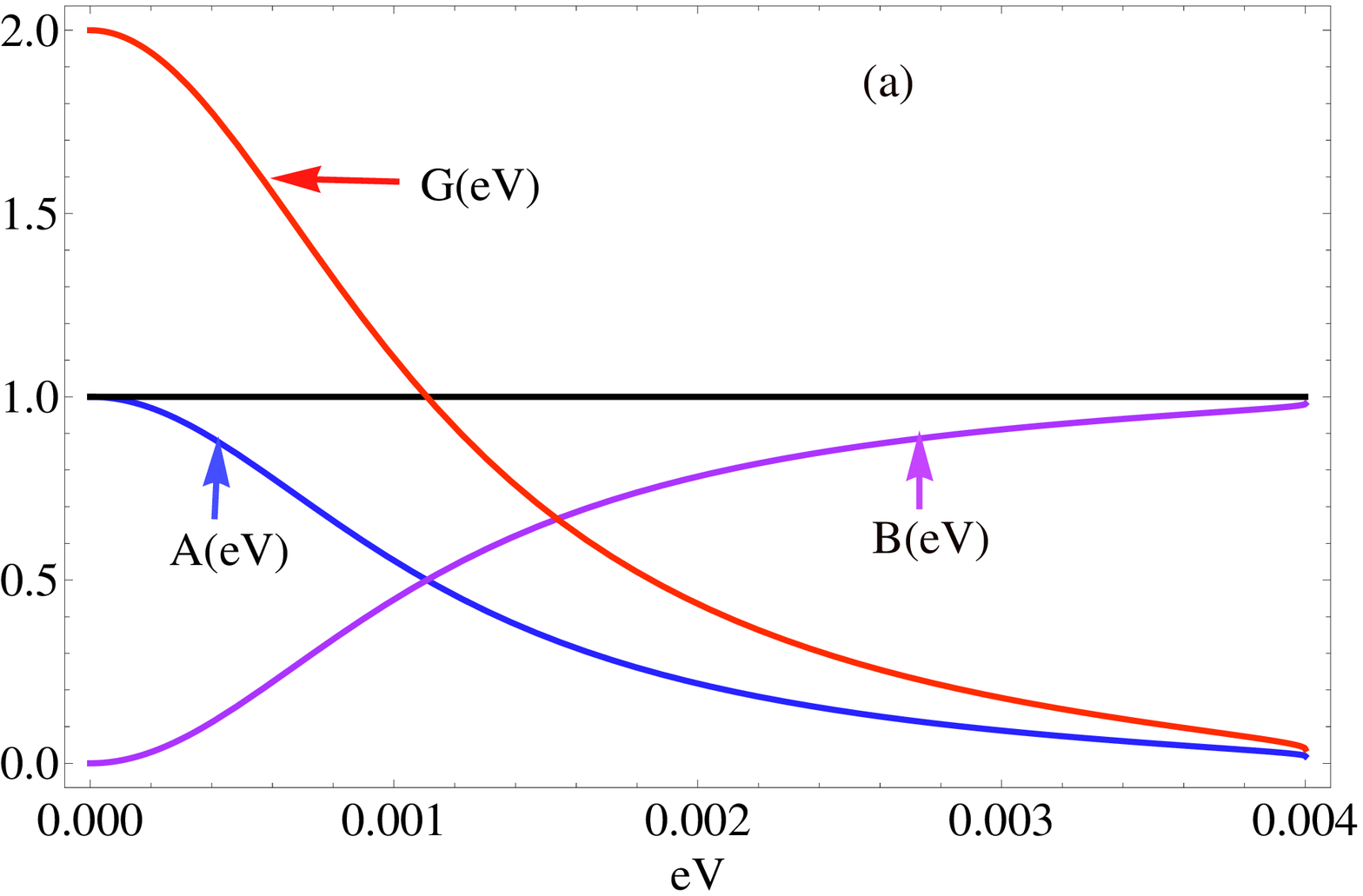}}
\subfigure{\includegraphics[width=4.0cm, height=3.0cm]{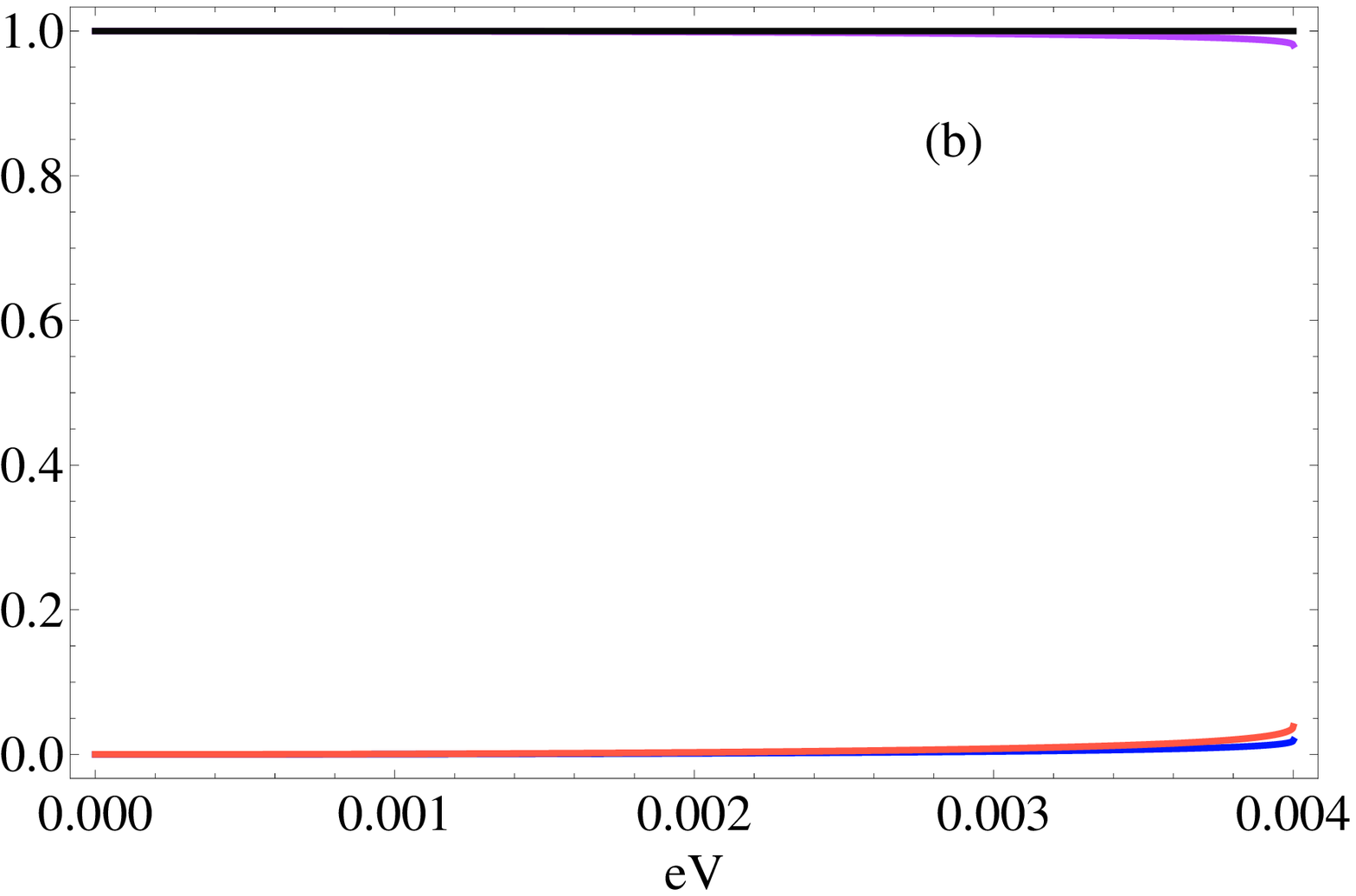}}
\subfigure{\includegraphics[width=4.0cm, height=3.0cm]{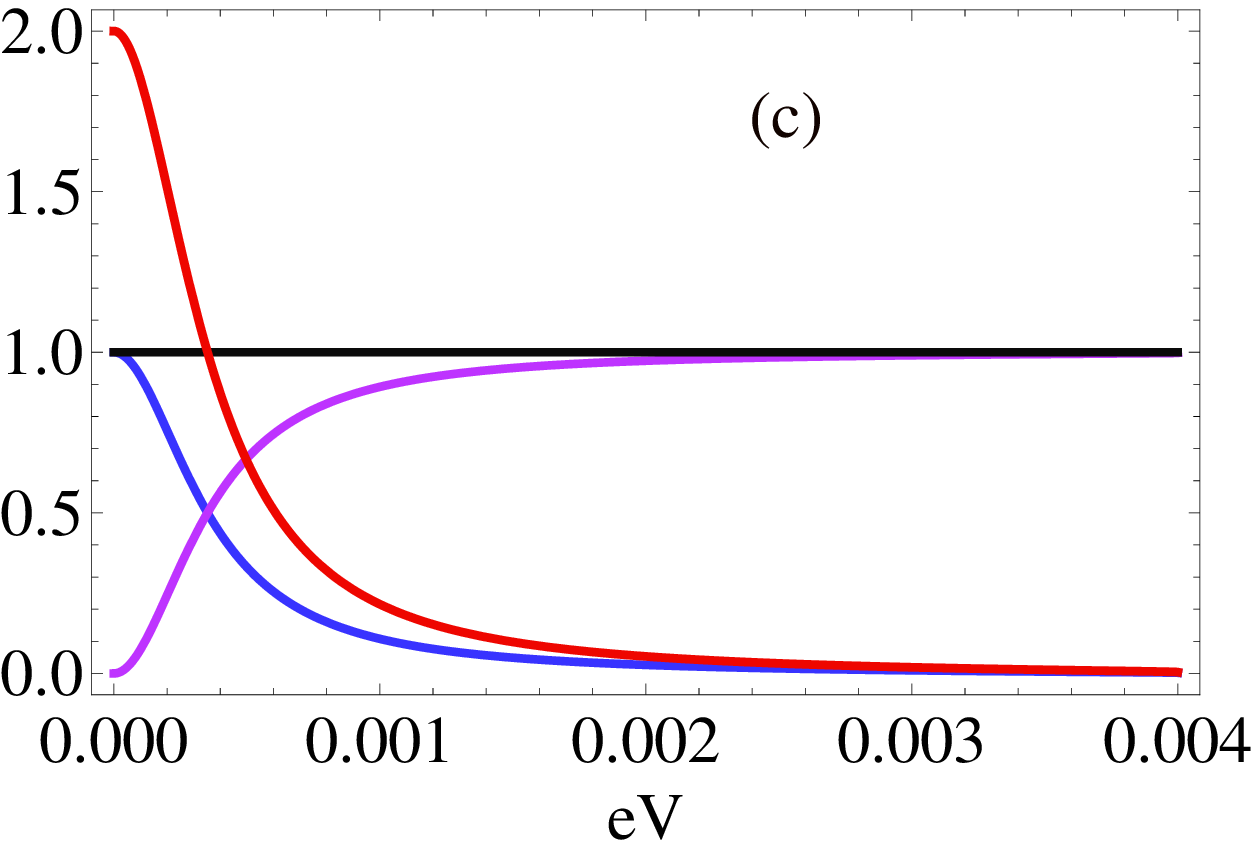}}
\subfigure{\includegraphics[width=4.0cm, height=3.0cm]{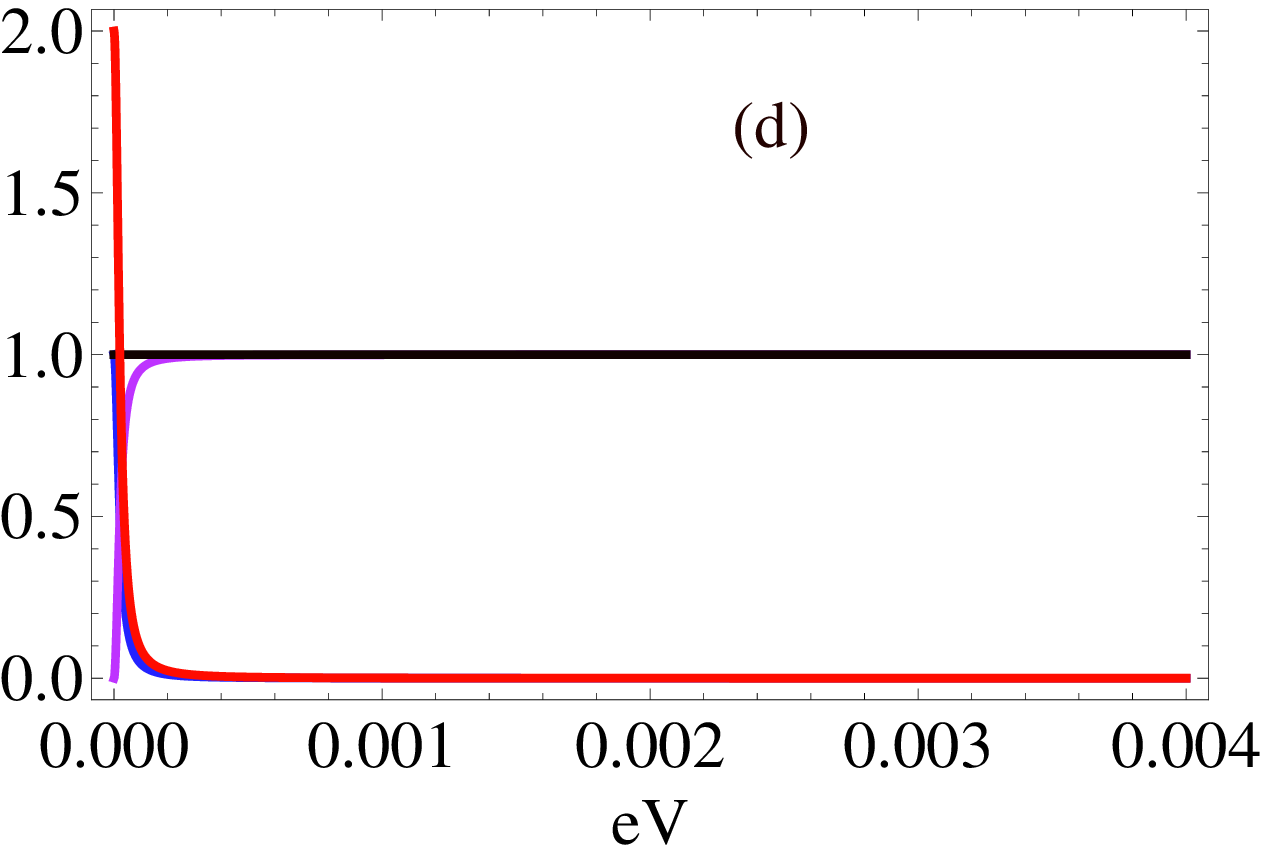}}
\subfigure{\includegraphics[width=4.0cm, height=3.0cm]{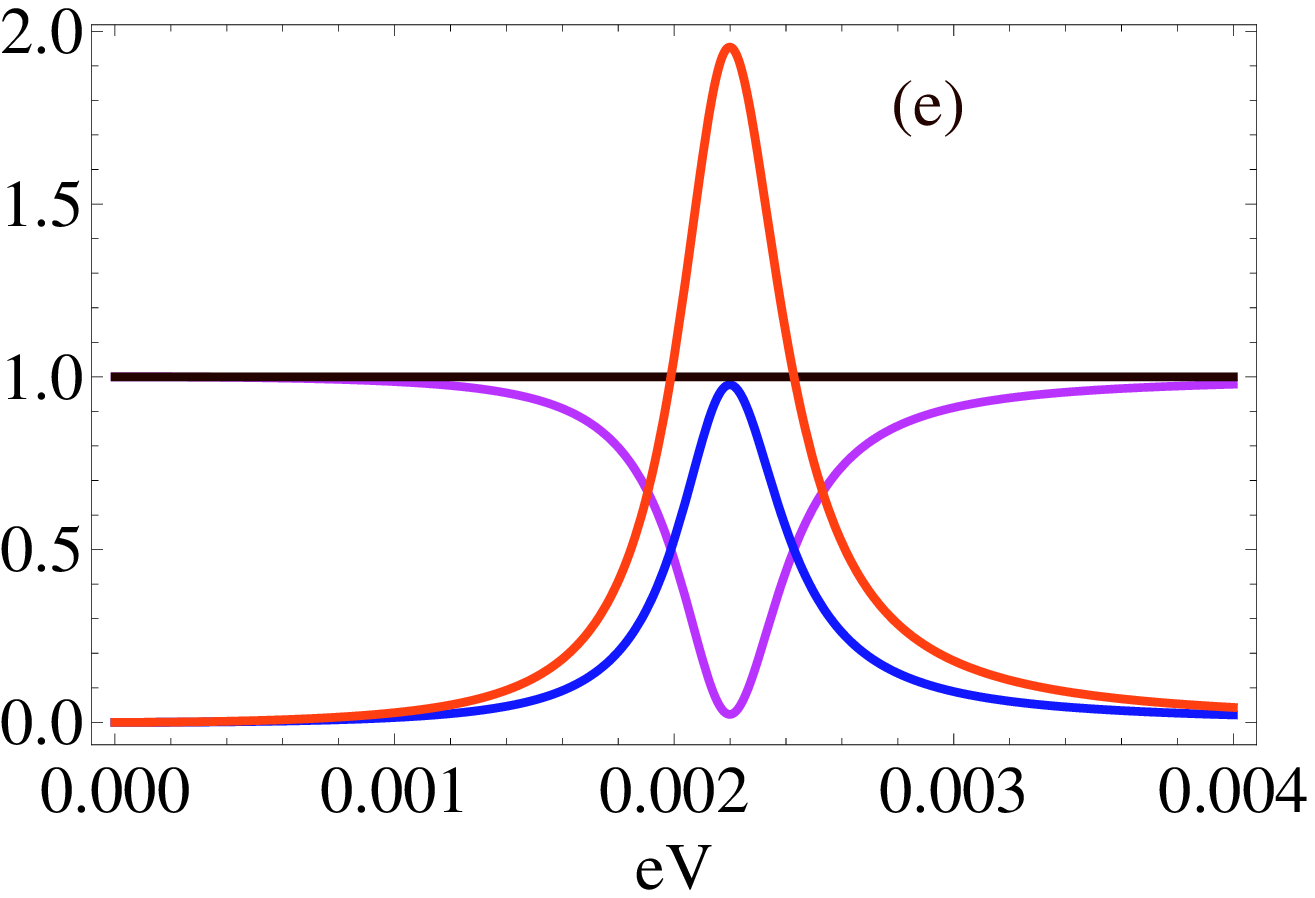}}
\subfigure{\includegraphics[width=4.0cm, height=3.0cm]{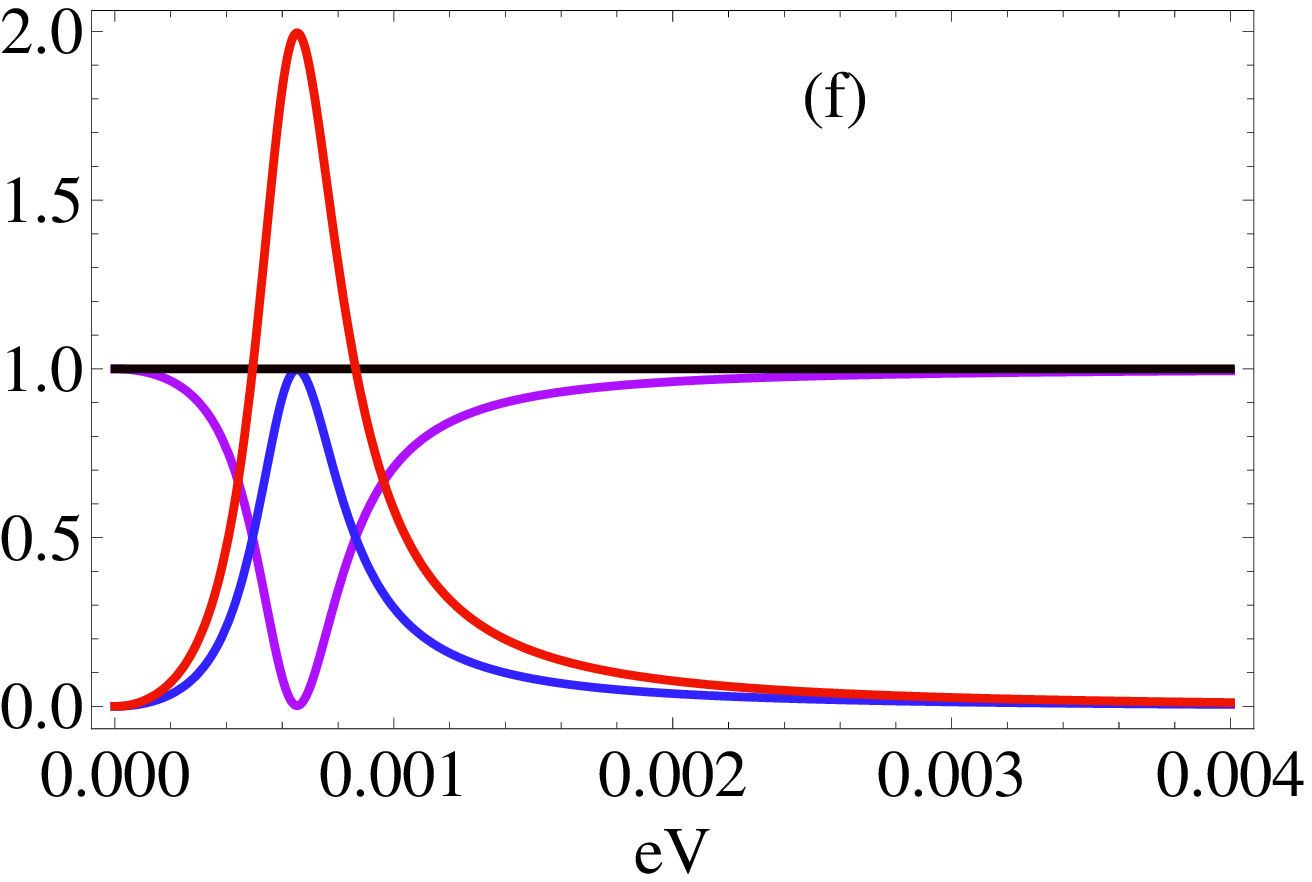}}
\caption{(color online)(a) Parameters: $L = \infty$, $Z = 0$,
$\mu_{n} = 1$, $\mu_{s} = 0.004$, $\Delta = 0.1$. $A(eV) =
|a(eV)|^{2} q_{h}/q_{e}$ is the equal-spin Andreev reflection
probability (blue line). $B(eV) = |b(eV)|^{2}$ is normal
reflection probability (purple line). $G(eV)/(e^{2}/h)$ is the
differential tunneling conductance (red line). The black line is
the normalization: $A(eV) + B(eV) = 1$. (b) $\mu_{s} = -0.004$,
(c) $Z = 2$, (d) $Z = 10$, (e) $L = 40$, $Z = 2$, (f)  $L = 60$,
$Z = 2$. The unmentioned parameters for (b)-(f) are the same as
(a).} \label{fig2}
\end{figure}

Following the BTK methods \cite{G. E. Blonder}, the two wave
functions (\ref{3}) and (\ref{5}) need to satisfy the boundary
conditions,
\begin{eqnarray}
&&\psi_{S}(x=L) = 0;\nonumber \\
&&\psi_{S}(x=0) = \psi_{N}(x=0); \nonumber\\
&&v_{s}\psi_{S}(x=0^{+}) - v_{n}\psi_{N}(x=0^{-}) = -iZ\sigma_{z} \psi_{N}(x=0), \nonumber\\
\label{6}
\end{eqnarray}
where $Z = 2 H$, $v_{s} = \partial H_{S}/\partial k$ and $v_{n} =
\partial H_{N}/\partial k$,  two $2 \times 2$ matrices, are the
velocity operators \cite{L. W. Molenkamp}. From Eq.(\ref{6}), we
can obtain $a(E)$ and $b(E)$ directly, and according to
Ref.\cite{G. E. Blonder}, the zero-temperature differential
tunneling conductance is given as
\begin{eqnarray}
G(eV) = \frac{e^{2}}{h} \left[1 + A(eV) - B(eV) \right], \label{7}
\end{eqnarray}
where $A(E) = |a(E)|^{2} q_{h}/q_{e}$, ($E=eV$), is the equal-spin
Andreev reflection probability, and $B(E) = |b(E)|^{2}$ is the
normal reflection probability. Here as we only consider in-gap
states, the waves expressed in Eq.(\ref{3}) do not carry current,
and $A(eV)$ and $B(eV)$ satisfy the normalization condition,
$i.e.$, $A(eV) + B(eV) = 1$. For different parameters $\mu_{s}$,
$L$ and $Z$ (we set $\mu_{n} = 1$ as unit), the results are shown
in Fig.\ref{fig2}.

Fig.\ref{fig2}(a) and Fig.\ref{fig2}(b) show the important
difference of differential tunneling conductance between $\mu_{s}
> 0$ (topologically non-trivial) and $\mu_{s} < 0$ (topologically
trivial). For $\mu_{s} > 0$, a quantized zero-bias conductance
peak of $2e^{2}/h$ stably exists (as shown in
Fig.\ref{fig2}(a)(c)(d)). A stable and quantized zero-bias
conductance peak is a manifestation of perfect equal-spin Andreev
reflection and indicates a stable topological phase. This result
is the same as the one obtained by an ``interface
electron-Majorana hopping" model \cite{K. T. Law, Karsten
Flensberg}, therefore, even we do not, in prior, assume the
existence of the Majorana bound states at the wire end, the
non-trivial topology of the $p$-wave superconductor manifest
itself in the tunneling spectroscopy obtained by matching the wave
functions of the two bulks directly. For $\mu_{s} < 0$, the in-gap
differential tunneling conductance almost vanishes. The important
difference indicates that, for $p$-wave superconductor, the
tunneling experiments can detect the topological quantum
transition at $\mu_{s} = 0$ effectively. Fig.\ref{fig2}(c) and
Fig.\ref{fig2}(d) show that increasing the interface scattering
potential only narrows the width of the peak,  neither changes the
quantized height (an analytical proof for it is given in the
Appendix) nor the position of the peak. The wire length has strong
impact on the conductance for $\mu_{s}>0$, as shown in
Fig.\ref{fig2}(e) and Fig.\ref{fig2}(f), for $L=40$ and $L=60$,
the conductance peaks appear at finite-bias voltage and the peak
will move toward to left (zero-bias voltage) with increasing $L$
(this agrees with the usual two end-Majoranas coupled picture).
The effect of the interface scattering potential for finite length
is similar to the one mentioned above. For sufficiently short $L$,
for example, $L < 10$, there is no conductance peak and the in-gap
differential tunneling conductance almost vanishes. For $\mu_{s} <
0$, the topologically trivial phase, the wire length $L$ and the
interface scattering potential $Z$ has little effect, the in-gap
differential tunneling conductance keeps very small.

Above we have restricted $\mu_{s}$ to be close to $\mu_{c}$, and
there is a big mismatch between $\mu_{n}$ and $\mu_{s}$. Canceling
this restriction, we find, for $\mu_{s}>0$, increasing $\mu_{s}$
greatly widens the width of the zero-bias peak of the in-gap
tunneling spectroscopy, however, the peak's quantization behavior
does not change and therefore the main physics does not change.
For $\mu_{s}<0$, decreasing $\mu_{s}$ has little effect to the
in-gap tunneling spectroscopy.

\subsection{$N-hS$ junction}
\label{subsec2b}

The one-dimensional $N-hS$ junction is shown in Fig.\ref{fig1}(b).
For $x < 0$, the Hamiltonian is a generalized $4 \times 4$ matrix
form of Eq.(\ref{4}). For $0 < x < L$, now the Hamiltonian is
given as \cite{Roman M. Lutchyn} (in momentum space, under
representation $\Psi^{\dag}_{k} = (c^{\dag}_{k,\uparrow},
c^{\dag}_{k,\downarrow}, c_{-k,\downarrow}, -c_{-k,\uparrow})$)
\begin{eqnarray}
H_{hS} = \xi_{k} \tau_{z} + B \sigma_{x} + \alpha k \sigma_{y}
\tau_{z} + \Delta \tau_{x}, \label{8}
\end{eqnarray}
where $\xi_{k} = k^{2}/2 - \tilde{\mu}_{s}$, $B$ is the in-plane
magnetic field along the wire, $\alpha$ is the spin-orbit coupling
strength, and $\Delta$ is the $s$-wave pair potential. $\vec{\tau}
= (\tau_{x}, \tau_{y}, \tau_{z})$ and $\vec{\sigma} =(\sigma_{x},
\sigma_{y}, \sigma_{z})$ are pauli matrices in particle-hole space
and spin space, respectively. The heterostructure described by
this Hamiltonian is just the one that is realized in the
experiment \cite{V. Mourik}. The quasiparticle energy spectrum is
given as
\begin{equation}
E = \sqrt{\xi_{k}^{2} + (\alpha k)^{2} + \Delta^{2} + B^{2} \pm 2
\sqrt{\xi_{k}^{2} \alpha^{2} k^{2} + B^{2} \xi_{k}^{2} + B^{2}
\Delta^{2}}}, \nonumber
\end{equation}
the energy gap is closed when the magnetic field $B$ reaches the
critical value $B_{c} = \sqrt{\tilde{\mu}_{s}^{2} + \Delta^{2}}$,
which separates $B < \sqrt{\tilde{\mu}_{s}^{2} + \Delta^{2}}$, the
topologically trivial phase, from $B > \sqrt{\tilde{\mu}_{s}^{2} +
\Delta^{2}}$, the topologically non-trivial phase. The
energy-momentum relation here is much more complicated than the
$p$-wave superconductor's. This complication makes us unable to
write down the analytical form of the wave function $\psi_{hS}(x)$
directly, and have to seek help from numerical tools.

As the Hamiltonian (\ref{8}) is a $4 \times 4$ matrix, the wave
function in the normal metal is a four-component vector, which
takes the form,
\begin{eqnarray}
\psi_{N}(x) = \left( \begin{array}{c}
              1 \\
              0 \\
              0 \\
              0
            \end{array} \right) e^{iq_{e}x} + \left( \begin{array}{c}
              b_{\uparrow} \\
              b_{\downarrow} \\
              0 \\
              0
            \end{array} \right) e^{-iq_{e}x} + \left( \begin{array}{c}
              0 \\
              0 \\
              a_{\downarrow} \\
              a_{\uparrow}
            \end{array} \right) e^{iq_{h}x}, \label{9}
\end{eqnarray}
where $b_{\uparrow}(E)$ denotes the normal reflection amplitude,
$b_{\downarrow}(E)$ denotes the spin-reversed reflection
amplitude, $a_{\uparrow}(E)$ denotes the equal-spin Andreev
reflection amplitude, and $a_{\downarrow}(E)$ denotes the
spin-reversed Andreev reflection amplitude. Now the boundary
conditions take the form
\begin{eqnarray}
&&\psi_{hS}(x=L) = 0; \nonumber \\
&&\psi_{hS}(x=0) = \psi_{N}(x=0); \nonumber\\
&&v_{hs} \psi_{hS}(x=0^{+}) - v_{n} \psi_{N}(x=0^{-}) = -iZ
\sigma_{0} \tau_{z} \psi_{N}(x=0), \nonumber\\
\label{10}
\end{eqnarray}
where $v_{hs} = \partial H_{hS}/\partial k$ is a $4 \times 4$
matrix, and $v_{n} = -i \partial_{x} \sigma_{0} \tau_{Z}$ is also
generalized to $4 \times 4$ matrix.

\begin{figure}
\subfigure{\includegraphics[width=4.0cm, height=3.0cm]{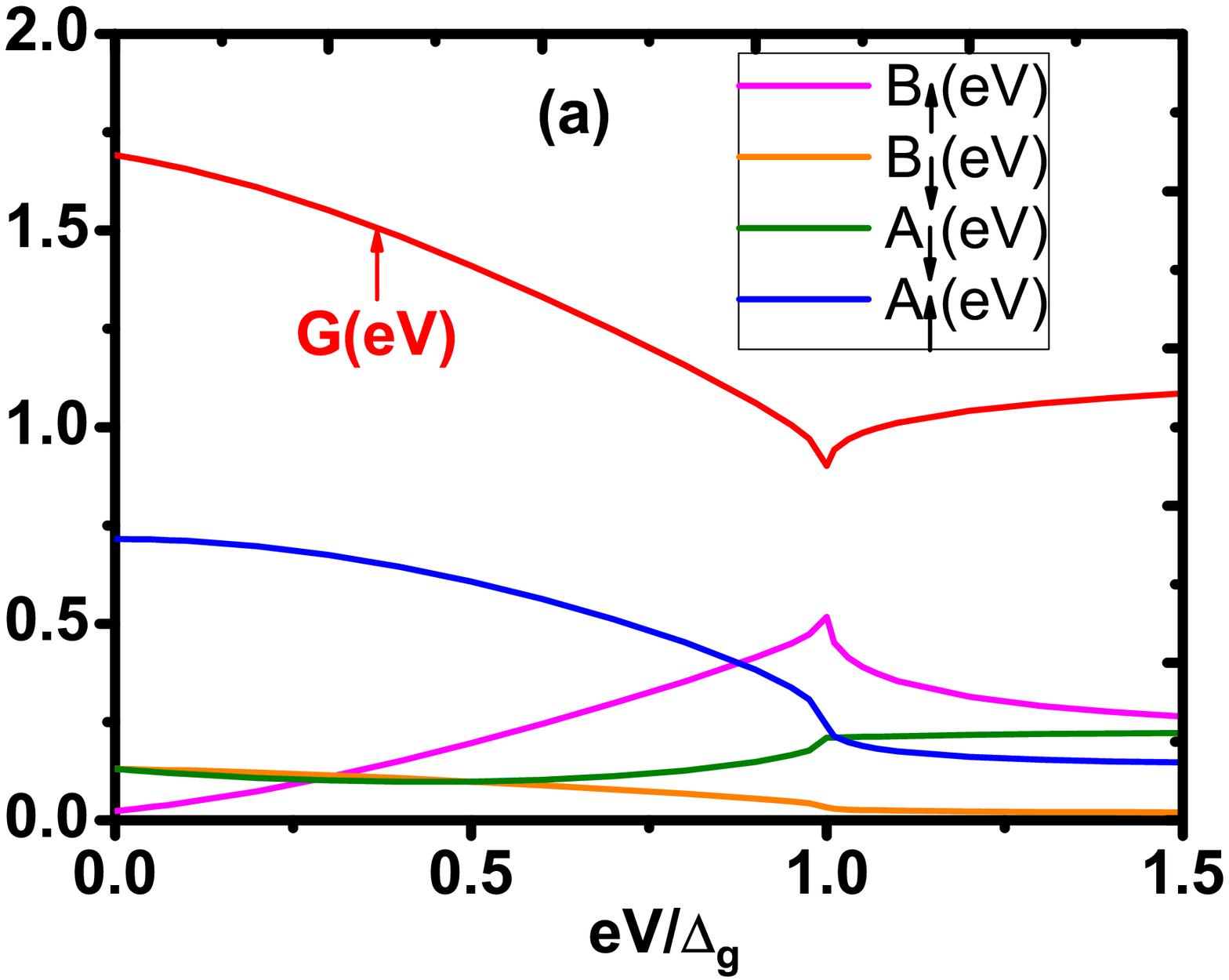}}
\subfigure{\includegraphics[width=4.0cm, height=3.0cm]{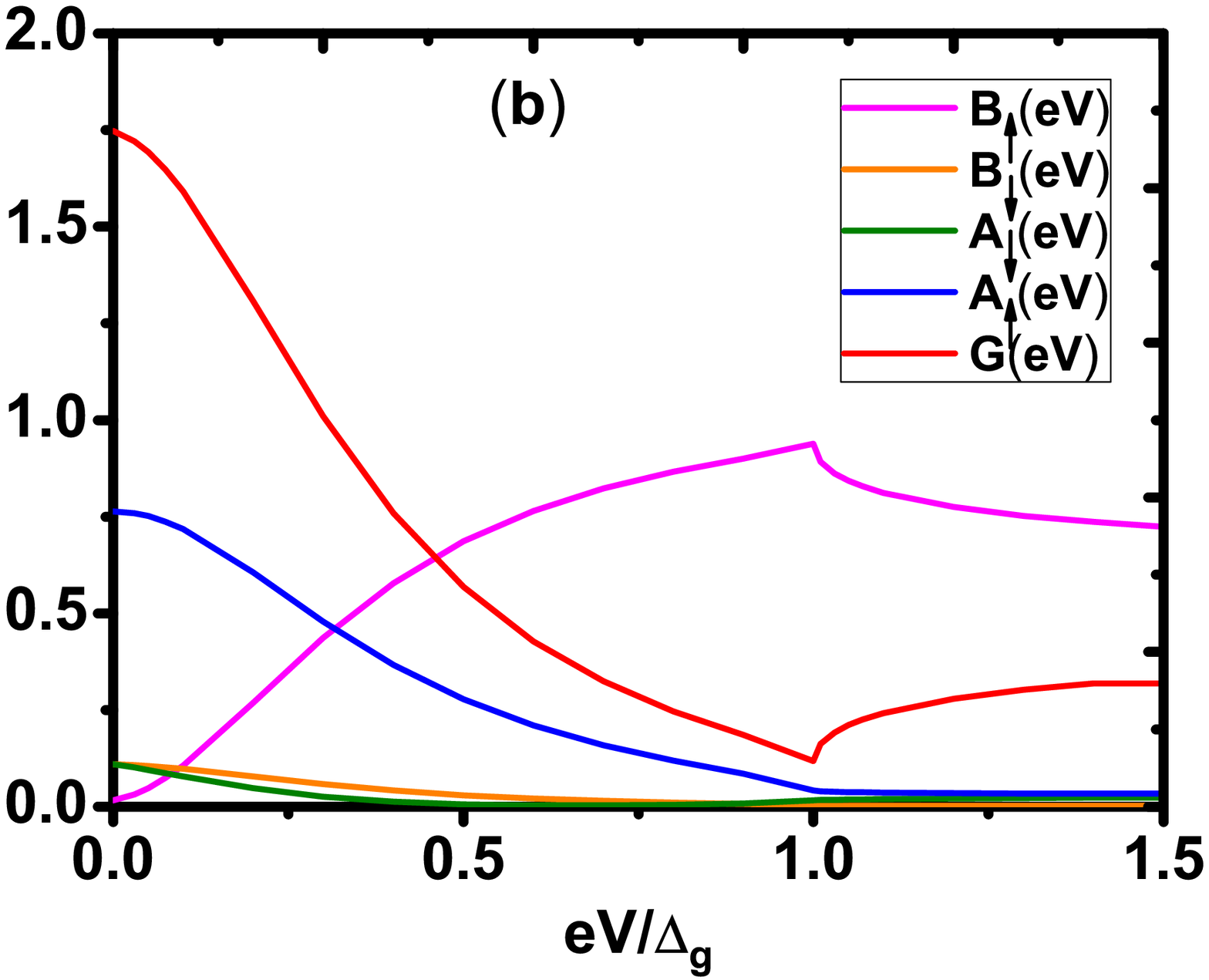}}
\subfigure{\includegraphics[width=4.0cm, height=3.0cm]{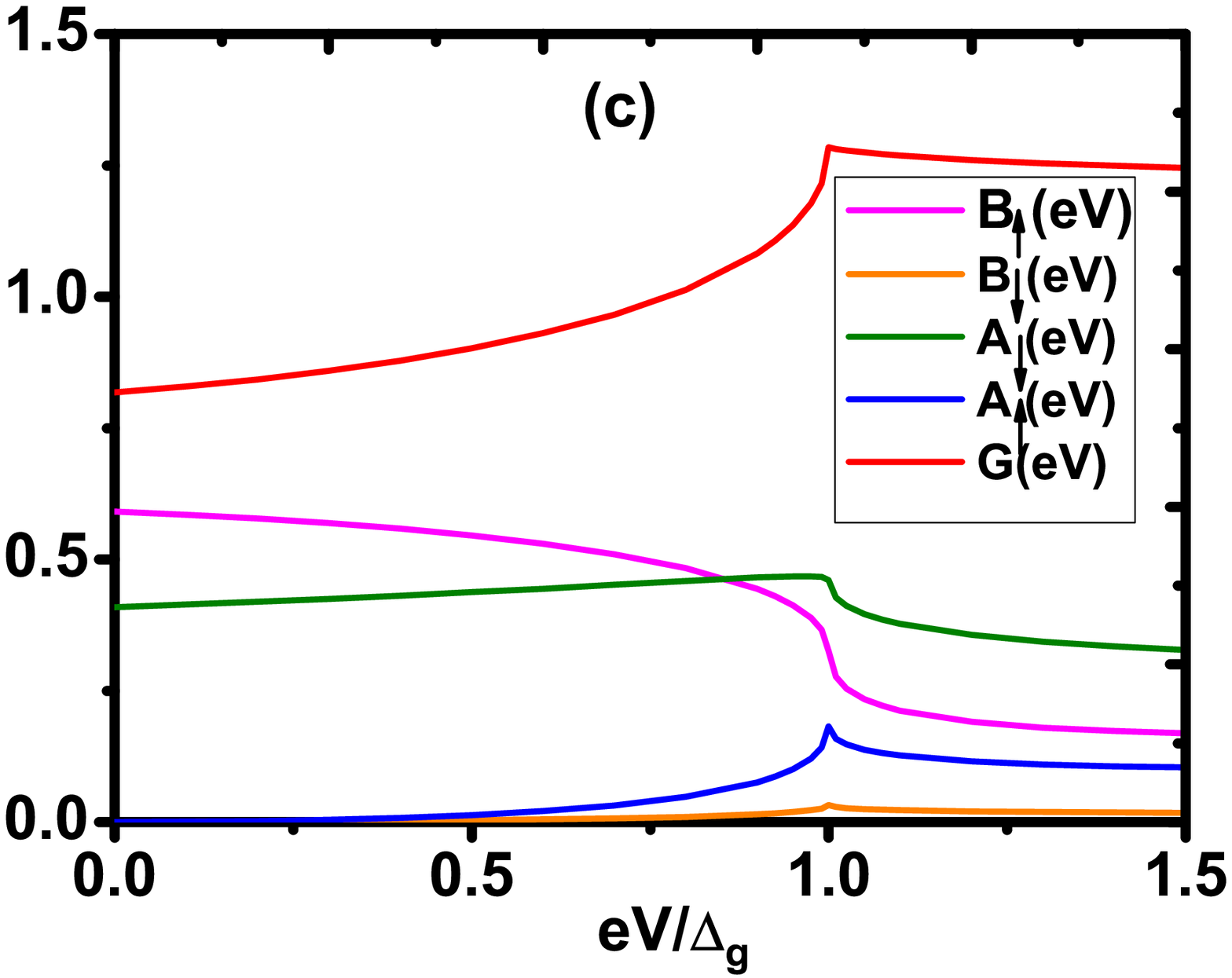}}
\subfigure{\includegraphics[width=4.0cm, height=3.0cm]{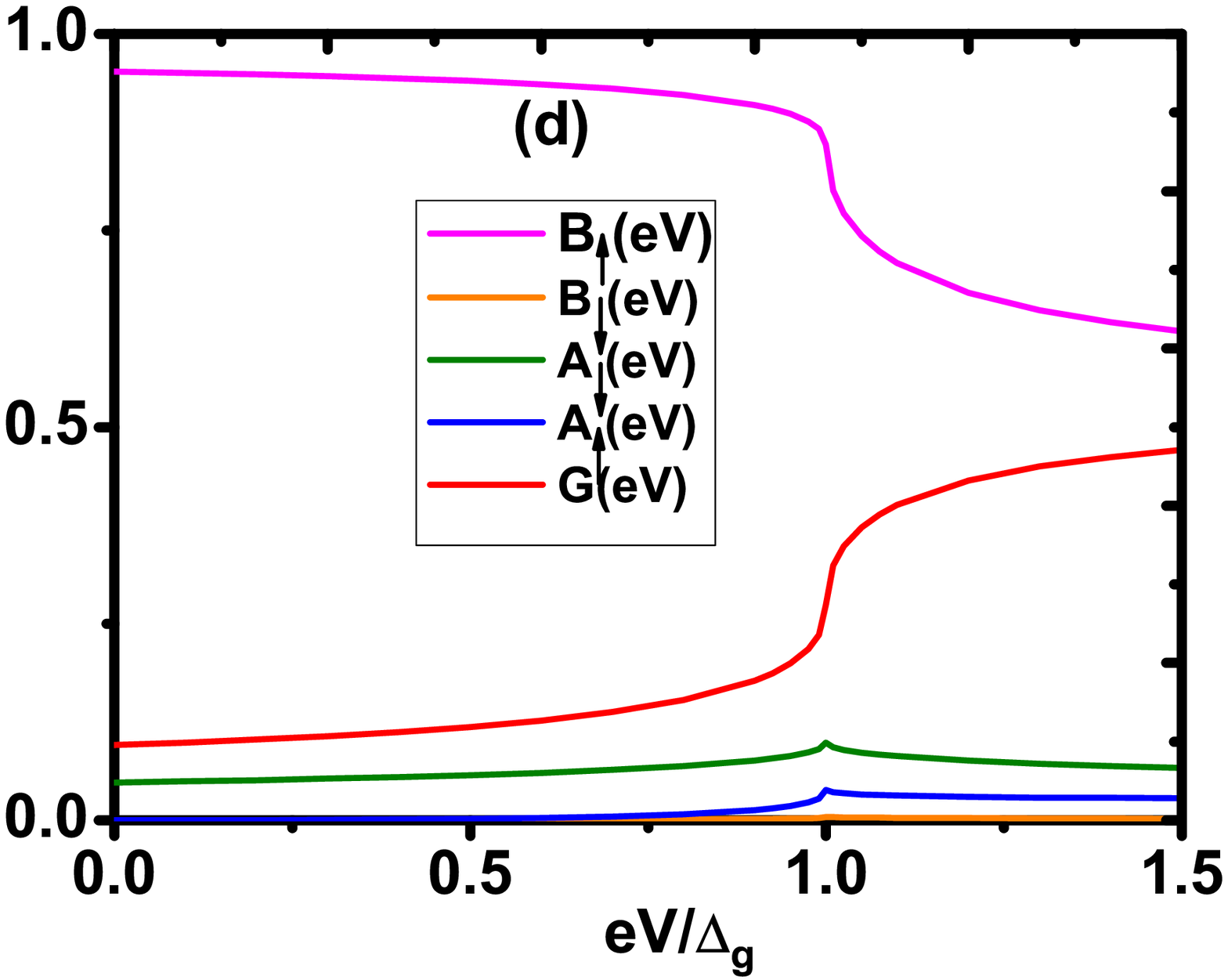}}
\caption{ (color online) Common parameters: $L=\infty$,
$\mu_{n}=1$, $\Delta=0.05$, $\Delta_{g}=0.01$. (a) $\alpha=0.5$,
$\mu_{s}=0$, $Z=0$, $B=0.06$. (b) $\alpha=0.5$, $\mu_{s}=0$,
$Z=2$, $B=0.06$. (c) $\alpha=0.5$, $\mu_{s}=0$, $Z=0$, $B=0.04$.
(d) $\alpha=0.5$, $\mu_{s}=0$, $Z=2$, $B=0.04$.
$A_{\uparrow}(eV)$, equal-spin Andreev reflection probability
(blue line), $A_{\downarrow}(eV)$, spin-reversed Andreev
reflection probability (olive line), $B_{\uparrow}(eV)$, normal
reflection probability (magenta line), $B_{\downarrow}(eV)$,
spin-reversed reflection probability (orange line),
$G(eV)/(e^{2}/h)$ is the differential tunneling conductance (red
line). In following figures, we adopt same color labeling.}
\label{fig3}
\end{figure}

Based on Eq.(\ref{10}), we can obtain $b_{\uparrow,\downarrow}(E)$
and $a_{\uparrow,\downarrow}(E)$, and the differential tunneling
conductance is given as
\begin{eqnarray}
G(eV) = \frac{e^{2}}{h} \left[1 + A_{\uparrow}(eV) +
A_{\downarrow}(eV) - B_{\uparrow}(eV) - B_{\downarrow}(eV)
\right], \nonumber \\
\label{11}
\end{eqnarray}
where $A_{\uparrow, \downarrow}(eV) = |a_{\uparrow,
\downarrow}(eV)|^{2} q_{h}/q_{e}$, and $B_{\uparrow,
\downarrow}(eV) = |b_{\uparrow, \downarrow}(eV)|^{2}$.
$A_{\uparrow, \downarrow}(eV)$ and $B_{\uparrow, \downarrow}(eV)$
in the gap region should satisfy the normalization condition:
$A_{\uparrow}(eV) + A_{\downarrow}(eV) + B_{\uparrow}(eV) +
B_{\downarrow}(eV) = 1$. For different parameters, the results are
shown in Fig.\ref{fig3}.

Fig.\ref{fig3}(a)(b) show when $B
> \sqrt{\tilde{\mu}_{s}^{2} + \Delta^{2}}$, the topological region, a
zero-bias conductance peak is formed. However, contrary to the
quantized zero-bias conductance peak of a $N-pS$ junction, here
the zero-bias peak is non-quantized and sensitive to parameters.
Increasing the interface scattering potential not only narrows the
width of the peak but also increases the height of the peak. From
Fig.\ref{fig3}(a)(b), we can also see the increase of the peak
height is due to a suppression of the spin-reversed Andreev
reflection and a simultaneous increase of the equal-spin Andreev
reflection by the interface scattering potential. However, with a
further increase of the interface scattering potential, this
corresponding increasing effect will finally be saturated, and the
zero-bias conductance peak still has a gap to the quantized value.

For comparison, Fig.\ref{fig3}(c)(d) show when $B <
\sqrt{\tilde{\mu}_{s}^{2} + \Delta^{2}}$, the normal phase, no
zero-bias peak appears, and by increasing the interface scattering
potential, the normal reflection is greatly enhanced and the
spin-reversed Andreev reflection and the conductance are greatly
reduced, which is a phenomenon familiar in $N-S$ junctions
\cite{G. E. Blonder}. Compared Fig.\ref{fig3}(a)(b) to
Fig.\ref{fig3}(c)(d), it is not hard to find that when the system
goes from the normal phase to the topological phase, the
probability of equal-spin Andreev reflection is greatly enhanced
(but still has a considerable gap to the perfect equal-spin
Andreev reflection) and the spin-reversed Andreev reflection
amplitude is greatly reduced, which indicates  the equal-spin
pairing ($p$-wave pairing) becomes much more favored in the
topological region than in the normal region.

\begin{figure}
\subfigure{\includegraphics[width=4.0cm, height=3.0cm]{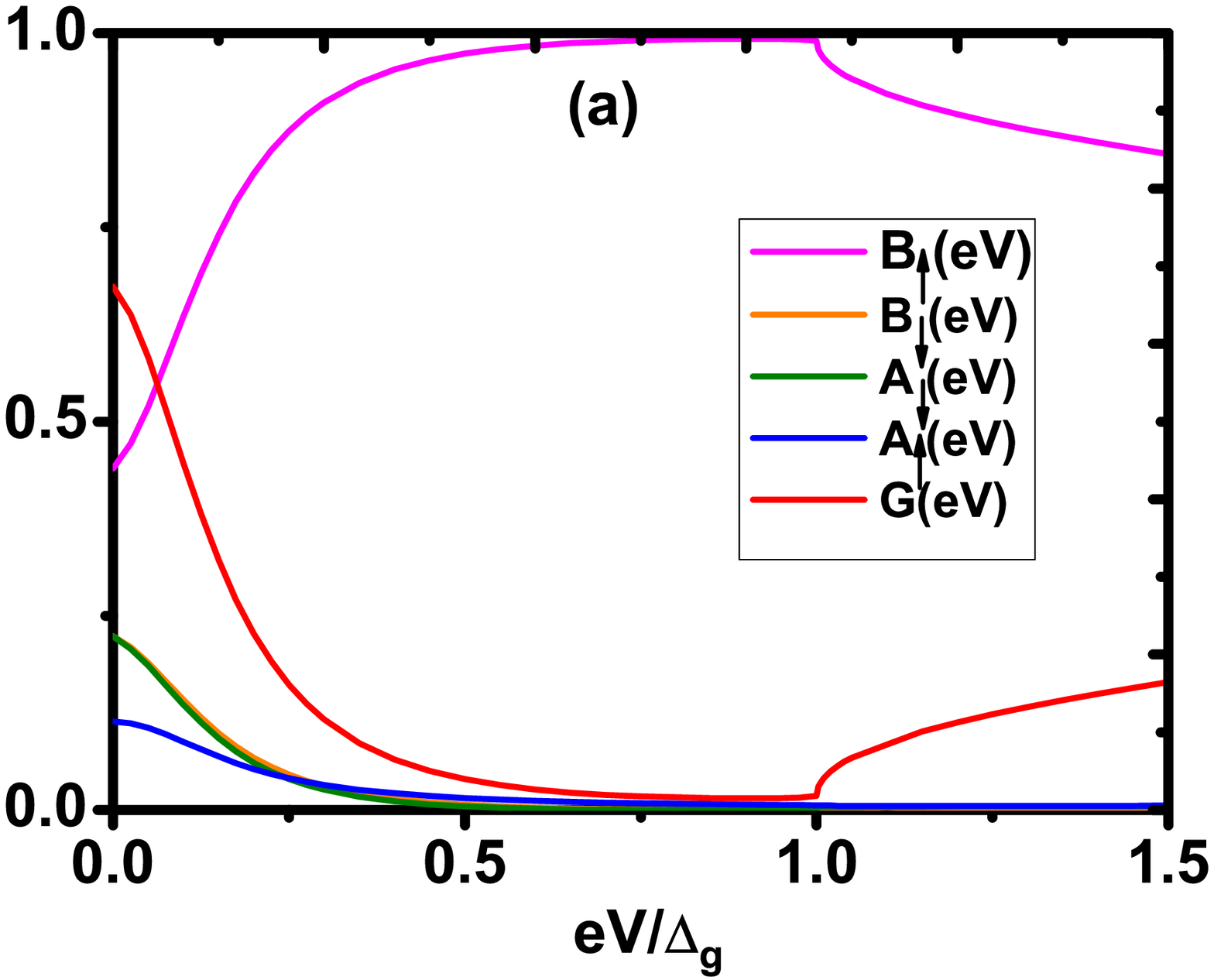}}
\subfigure{\includegraphics[width=4.0cm, height=3.0cm]{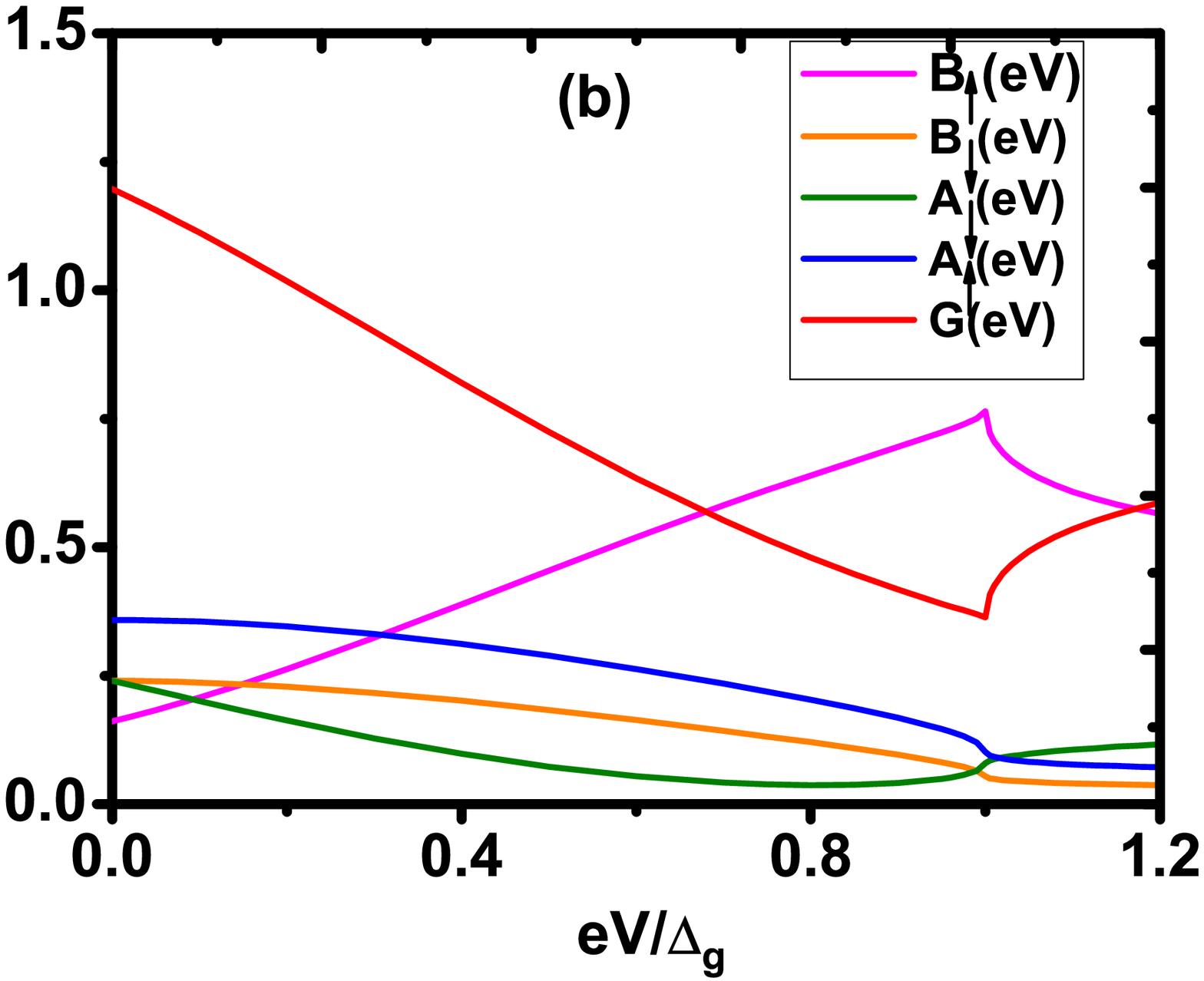}}
\subfigure{\includegraphics[width=4.0cm, height=3.0cm]{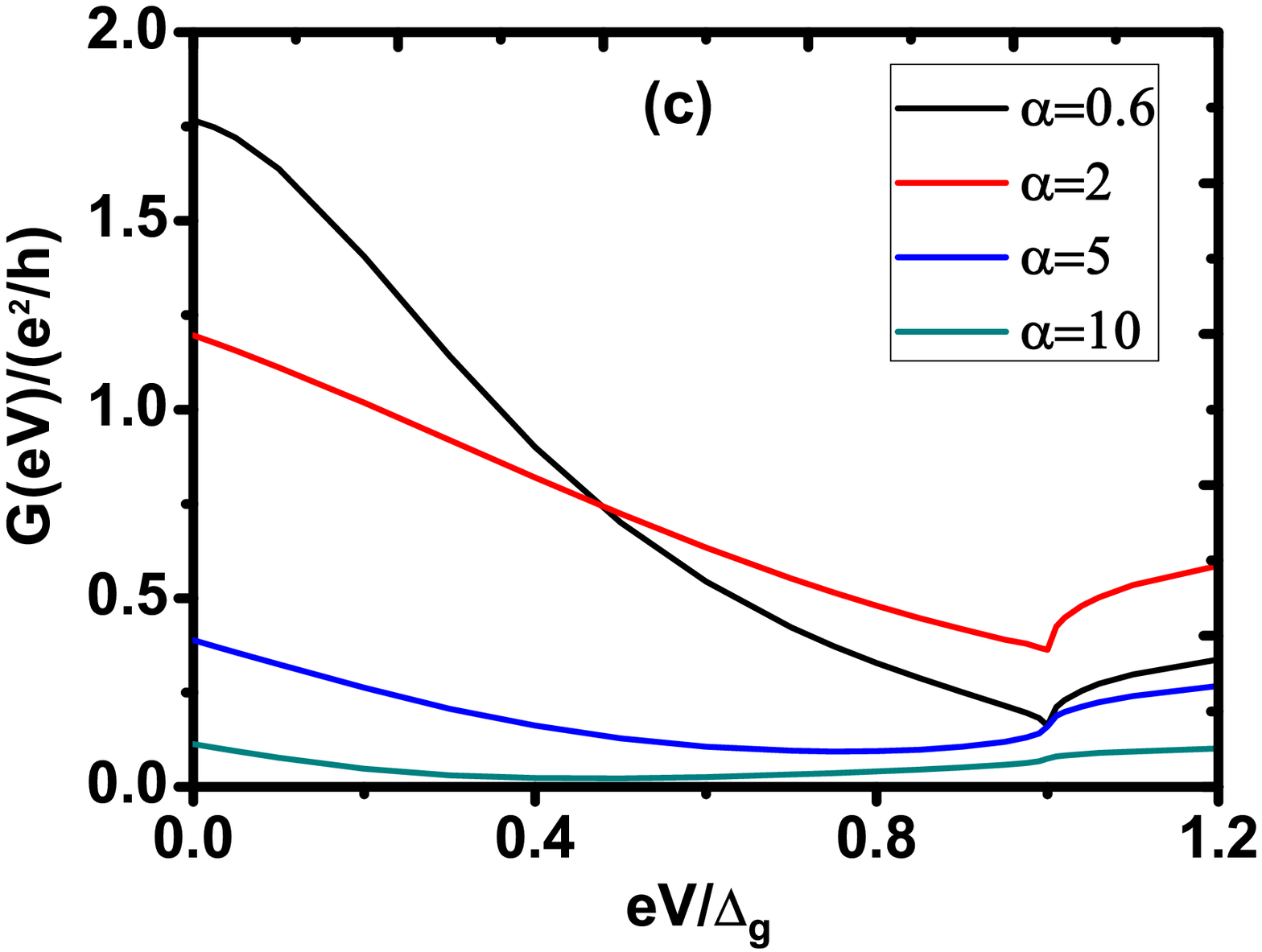}}
\subfigure{\includegraphics[width=4.0cm, height=3.0cm]{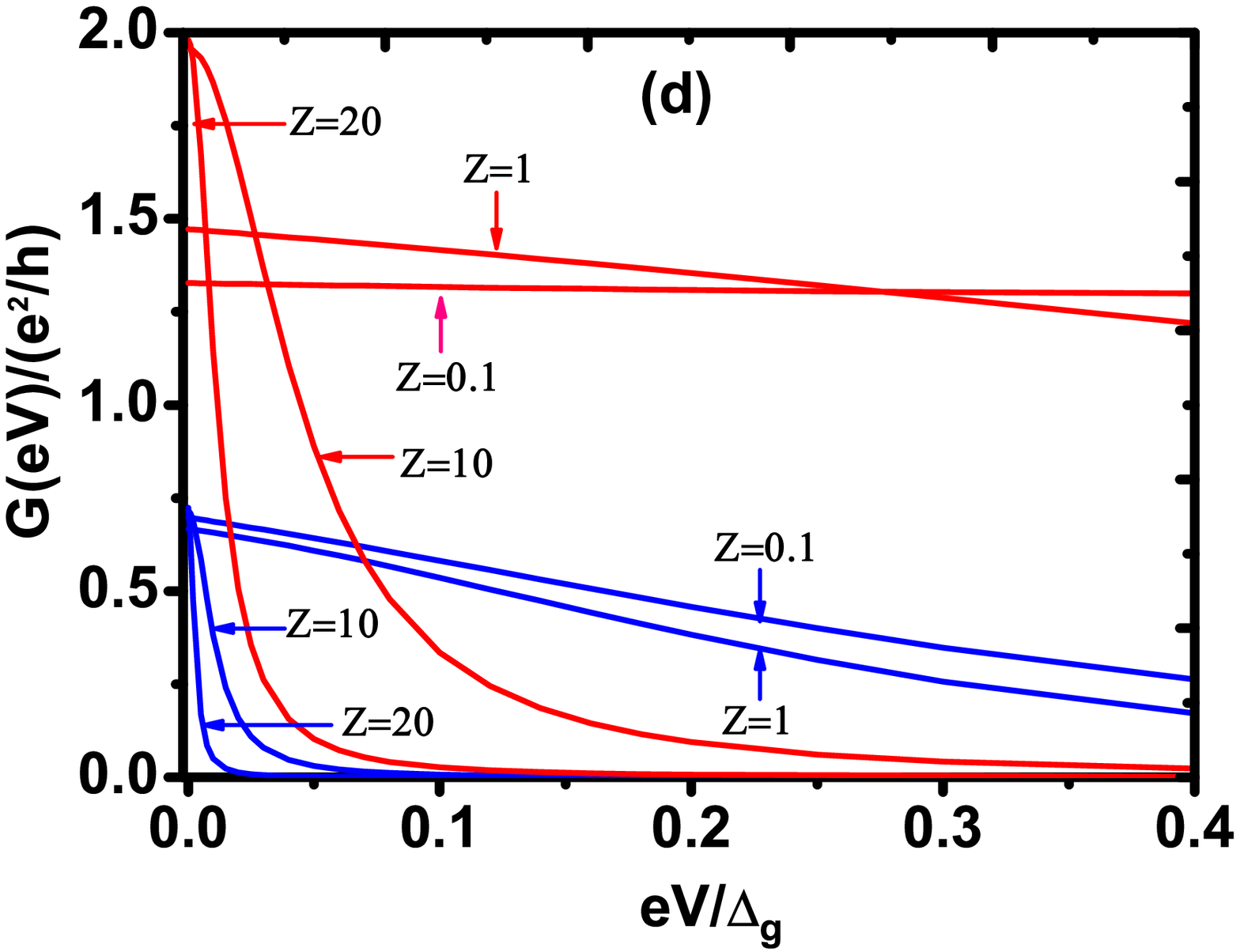}}
\caption{ (color online) Common parameters: $L=\infty$,
$\mu_{n}=1$, $\Delta=0.05$, $\mu_{s}=0$, $B=0.06$,
$\Delta_{g}=0.01$. (a) $\alpha=0.1$, $Z=2$, (b) $\alpha=2$, $Z=2$,
(c)$Z=2$, (d) red lines correspond to $\alpha=1$, blue lines
correspond to $\alpha=0.1$.} \label{fig4}
\end{figure}

Spin-orbit coupling also has strong impact on the tunneling
spectroscopy. As shown in Fig.\ref{fig4}(a), when decreasing the
spin-orbit coupling and fixing other parameters, the height of the
zero-bias peak and the probability of equal-spin Andreev
reflection are significantly reduced. A further inspection shows
that once the spin-orbit coupling decreases to a critical value
(named as $\alpha_{g}$) not only the peak height keeps decreasing
to a smaller value but also the induced gap $\Delta_{g}$ begins to
depend on spin-orbit coupling (when $\alpha < \alpha_{g}$, we say
the spin-orbit coupling is weak), with a dependence that it
monotonically decreases with decreasing spin-orbit coupling. When
the spin-orbit coupling is decreased to zero, the induced gap gets
closed, the system turns to be gapless and the zero-bias
conductance peak disappears, which indicates a breakdown of the
topological criterion

As above, we show that decreasing the spin-orbit coupling (from
the Fig.\ref{fig3}'s parameter, $\alpha = 0.5$) lowers the peak,
we find that increasing the spin-orbit coupling does not
correspond to a monotonic increase of the peak height. As shown in
Fig.\ref{fig4}(b)(c), the zero-bias conductance peak first
increases and then decreases with increasing spin-orbit coupling,
the optimal spin-orbit coupling $\alpha_{c}$ under the parameters
given in Fig.\ref{fig4}(c) is about $0.6$.  In the following, when
$\alpha_{g} < \alpha < \alpha_{c}$, we say the spin-orbit coupling
is in the intermediate region, and when $\alpha > \alpha_{c}$, we
say the spin-orbit coupling is strong. From Fig.\ref{fig4}(c), we
also find when $\alpha$ goes beyond $\alpha_{c}$, a larger
$\alpha$ corresponds to a lower peak and the reduction effect due
to the increase of spin-orbit coupling is significant. However, we
also find, almost simultaneously when $\alpha$ goes beyond
$\alpha_{c}$, the saturation effect of the interface scattering
potential for weak spin-orbit coupling is absent, and a stronger
interface potential will induce a higher peak, as shown in
Fig.\ref{fig4}(d). When the interface scattering potential goes to
infinity, the peak goes to the quantized value, $i.e.$,
$2e^{2}/h$, and the width of the peak goes to zero. A quantized
peak located at zero-bias voltage with vanishing width is
apparently a manifestation of the Majorana end states.

\begin{figure}
\subfigure{\includegraphics[width=4.0cm, height=3.0cm]{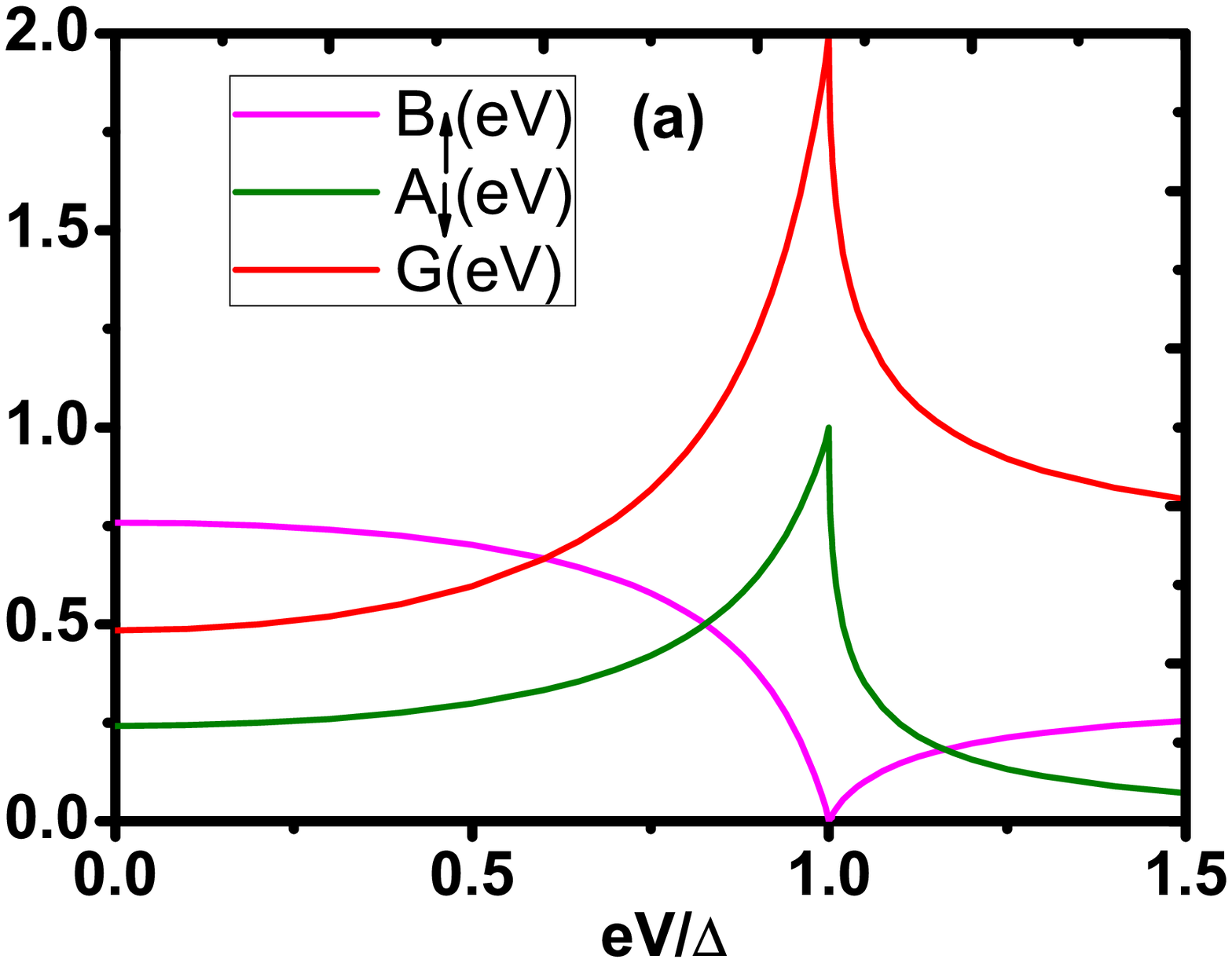}}
\subfigure{\includegraphics[width=4.0cm, height=3.0cm]{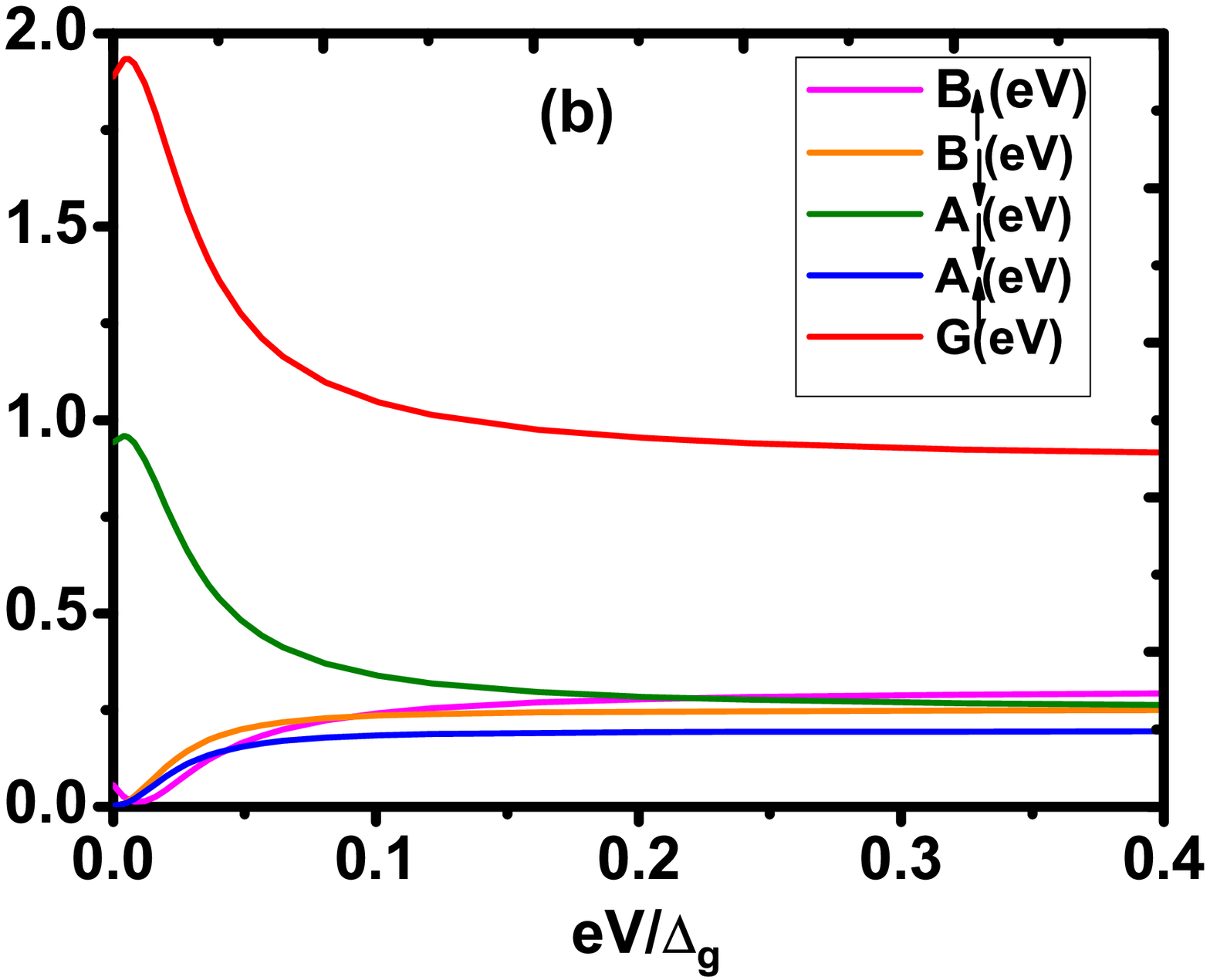}}
\subfigure{\includegraphics[width=4.0cm, height=3.0cm]{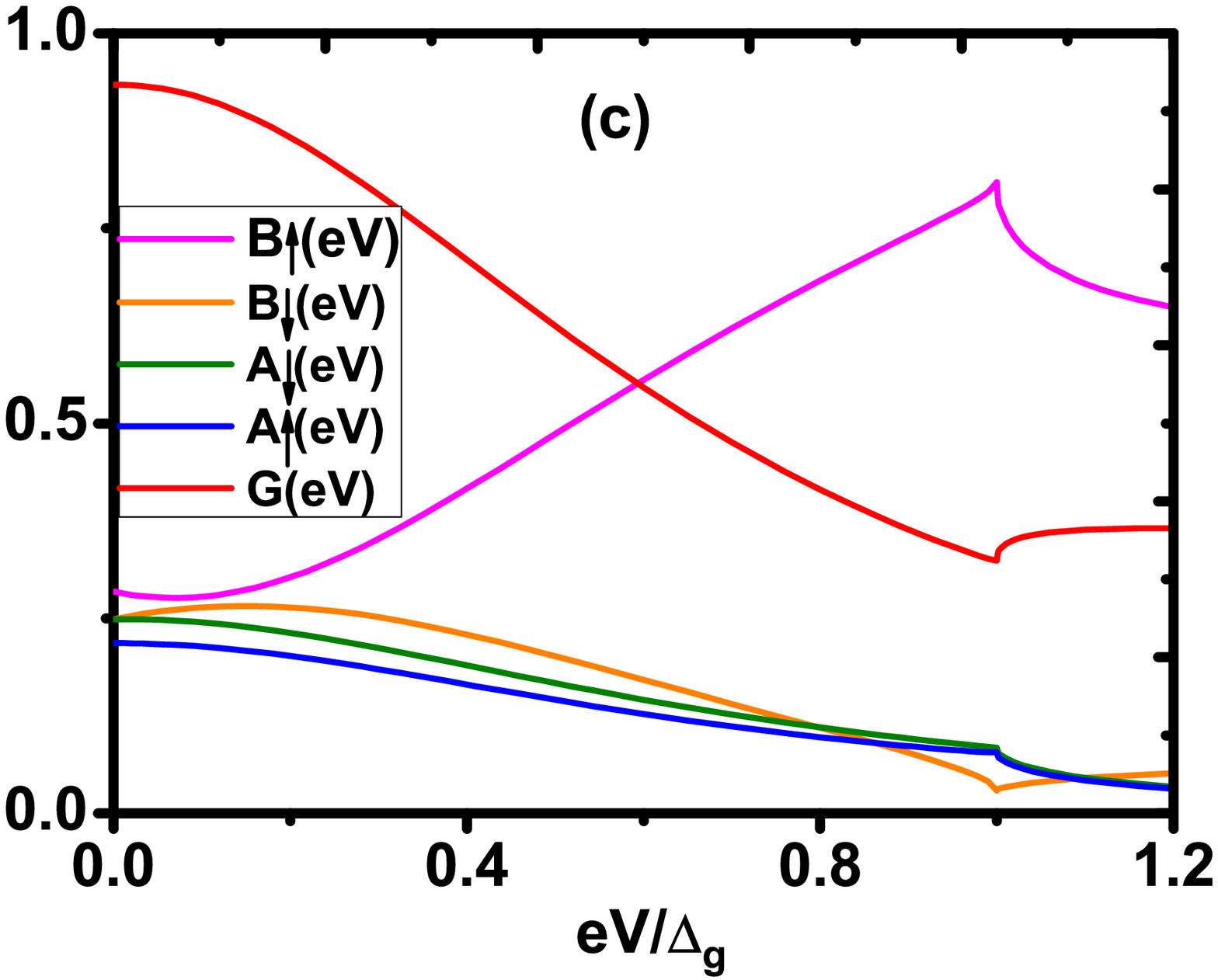}}
\subfigure{\includegraphics[width=4.0cm, height=3.0cm]{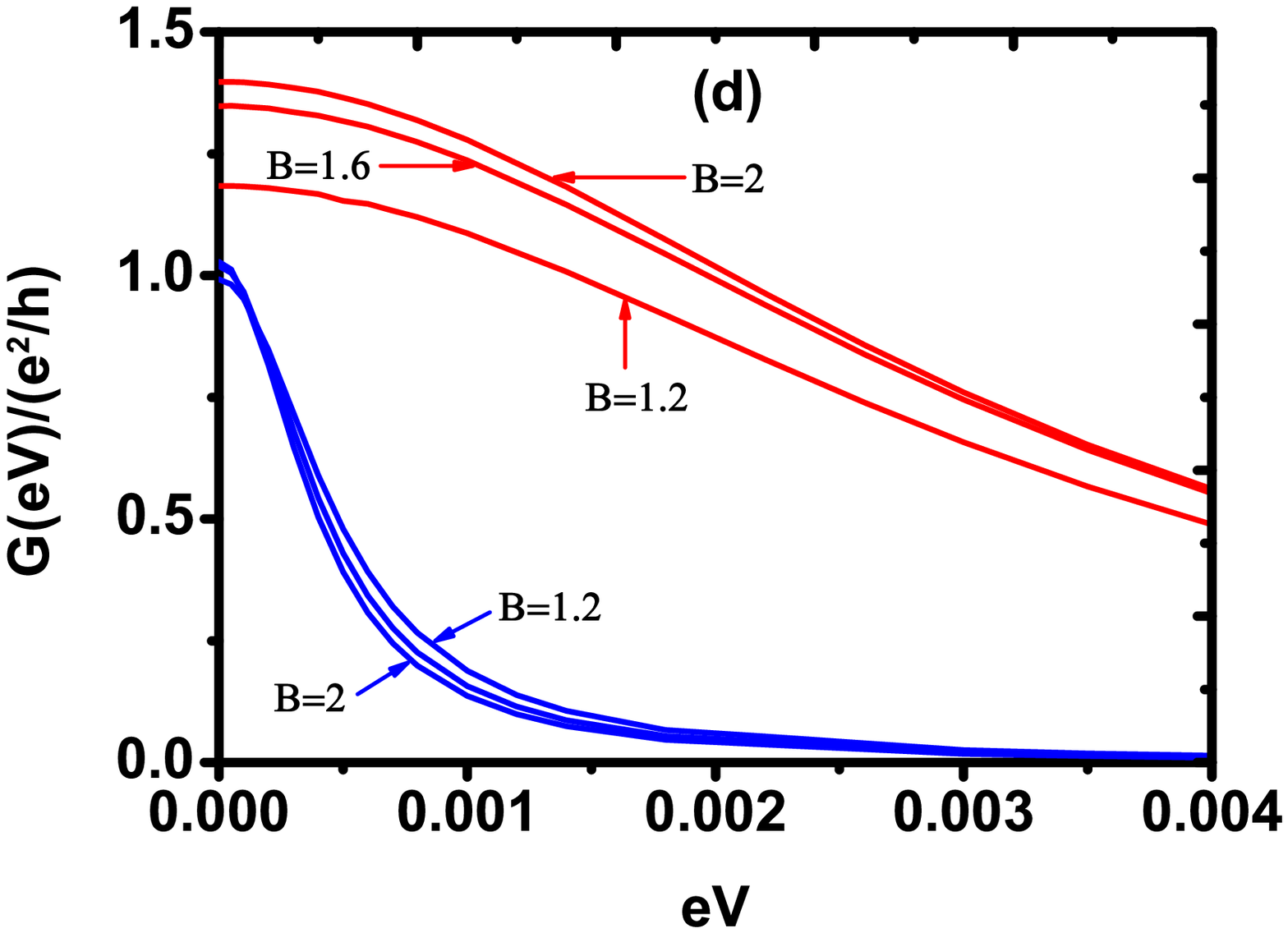}}
\caption{ (color online) Common parameters: $L=\infty$,
$\mu_{n}=1$, $\mu_{s}=1$, $\Delta=0.05$, $\alpha=0.1$, $Z=2$. (a)
$B=0$, (b) $B=1$, $\Delta_{g}=0.00025$, (c) $B=1.1$,
$\Delta_{g}=0.0092$, (d) red lines correspond to $\alpha=1$, blue
lines correspond to $\alpha=0.1$, the middle blue line corresponds
to $B=1.6$.}
\label{fig5}
\end{figure}

Fig.\ref{fig5}(a) shows that when the chemical potentials are not
mismatch between the normal lead and the heterostructure
superconductor and the magnetic field is absent, there exists a
sharp peak corresponding to a resonant spin-reversed Andreev
reflection at the induced-gap boundary, similar to the $N-S$
junction \cite{G. E. Blonder}. A fact that needs to notice is, in
the absence of magnetic field, there are no spin-reversed normal
reflection and equal-spin Andreev reflection even with strong
spin-orbit coupling. This indicates that the magnetic field is a
necessity to induce the equal-spin pairing. As shown in
Fig.\ref{fig5}(b), increasing the magnetic field will drive the
peak toward to left (zero-bias voltage) and the finite magnetic
field also drives the peak away from the induced-gap boundary.
However, when there is a big mismatch between the chemicals, which
is usually needed to guarantee that the magnetic filed satisfying
the topological criterion is still not large enough to break down
the superconductivity, such an interesting phenomenon is absent
(no finite-bias peak appears in Fig.\ref{fig3}(c)(d)). Once the
magnetic field reaches the critical value $B_{c} =
\sqrt{\tilde{\mu}_{s}^{2} + \Delta^{2}}$, a zero-bias conductance
peak is formed, however, with the height reduced a lot, as shown
in Fig.\ref{fig5}(c). A sudden reduction of the peak height maybe
imply the zero-bias conductance peak and the finite-bias peak are
due to different origins. For weak or intermediate spin-orbit
coupling, further increasing the magnetic field has little effect
on the zero-bias conductance peak, however, when the spin-orbit
coupling is strong enough, the peak height monotonically increases
to a parameter-dependent saturation value with increasing magnetic
field (here we do not consider the breakdown of superconductivity
due to a strong magnetic field), as shown in Fig.\ref{fig5}(d). A
further study in the stronger spin-orbit coupling region shows
that the absence of chemical potential mismatch makes the
zero-bias conductance peak approaching to the quantized value even
more difficult.

\begin{figure}
\subfigure{\includegraphics[width=4.0cm, height=3.0cm]{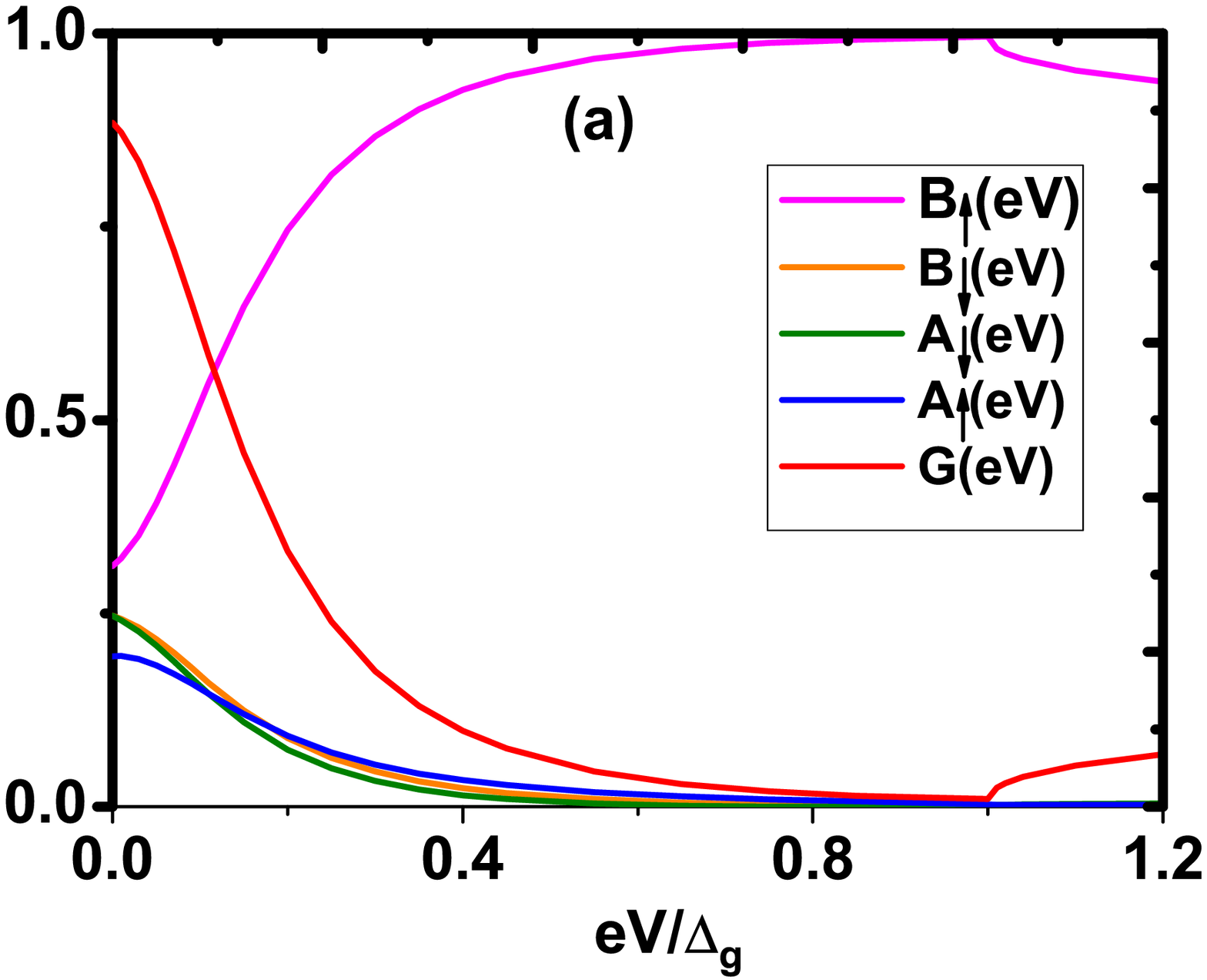}}
\subfigure{\includegraphics[width=4.0cm, height=3.0cm]{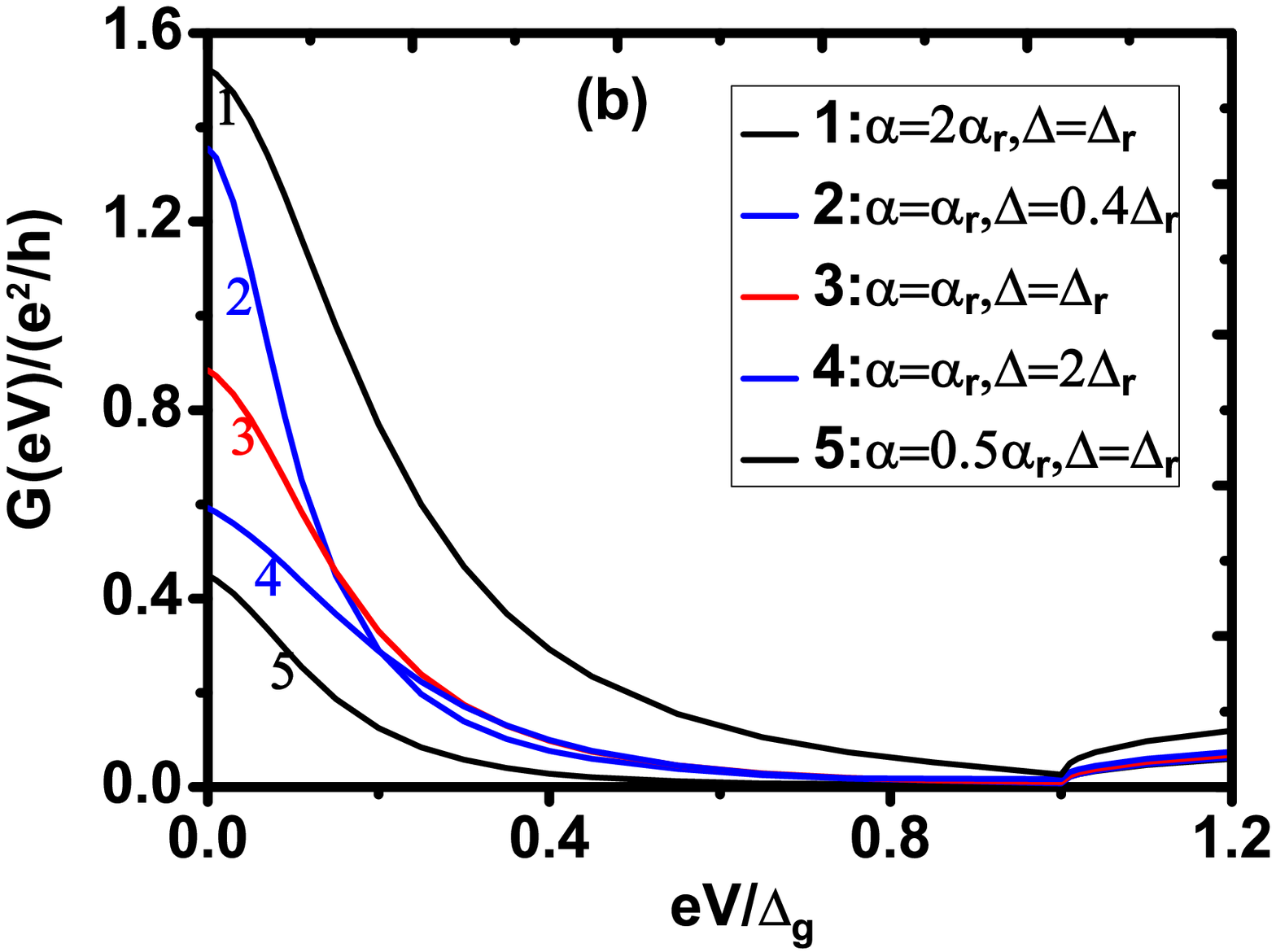}}
\subfigure{\includegraphics[width=4.0cm, height=3.0cm]{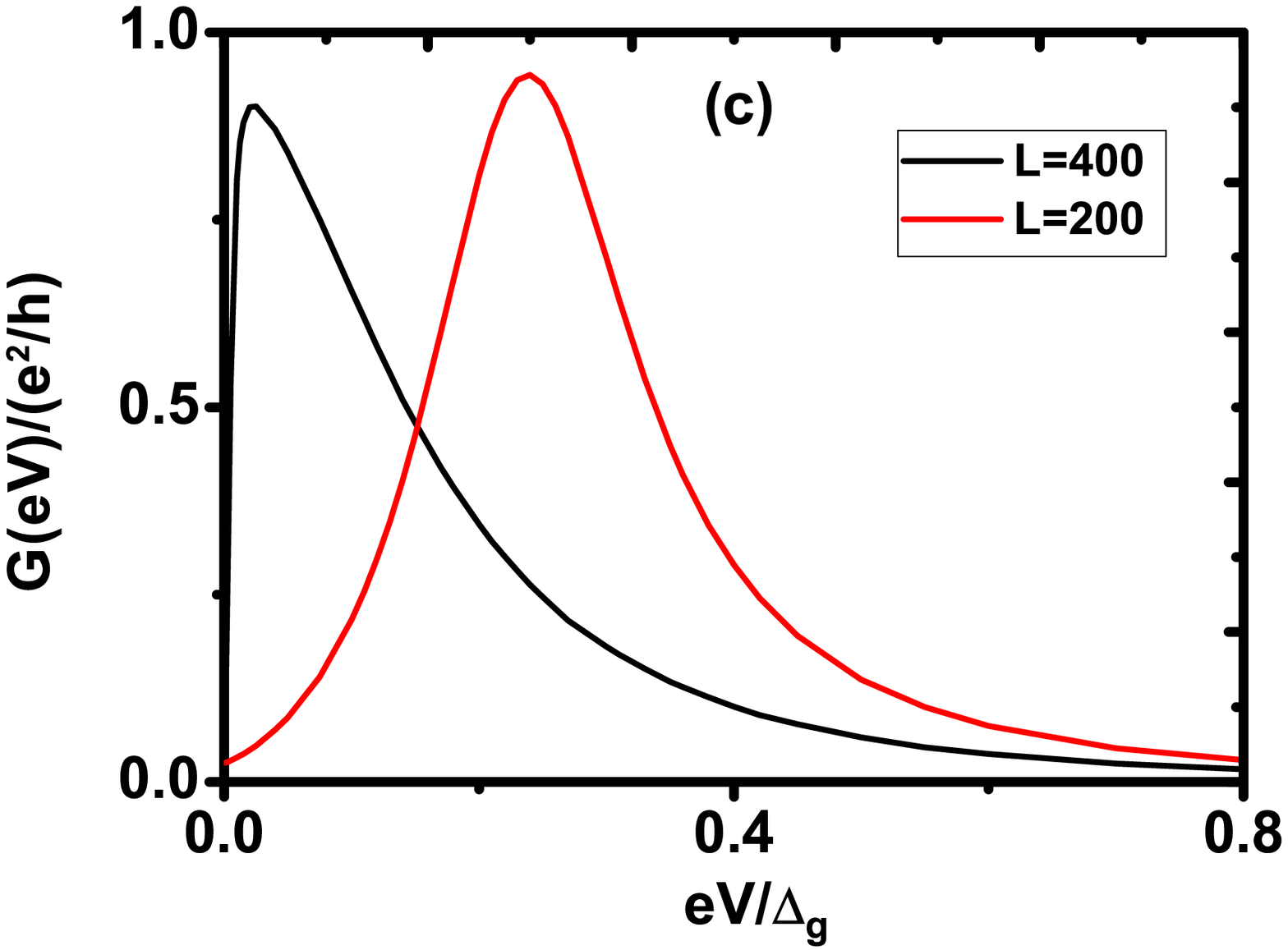}}
\subfigure{\includegraphics[width=4.0cm, height=3.0cm]{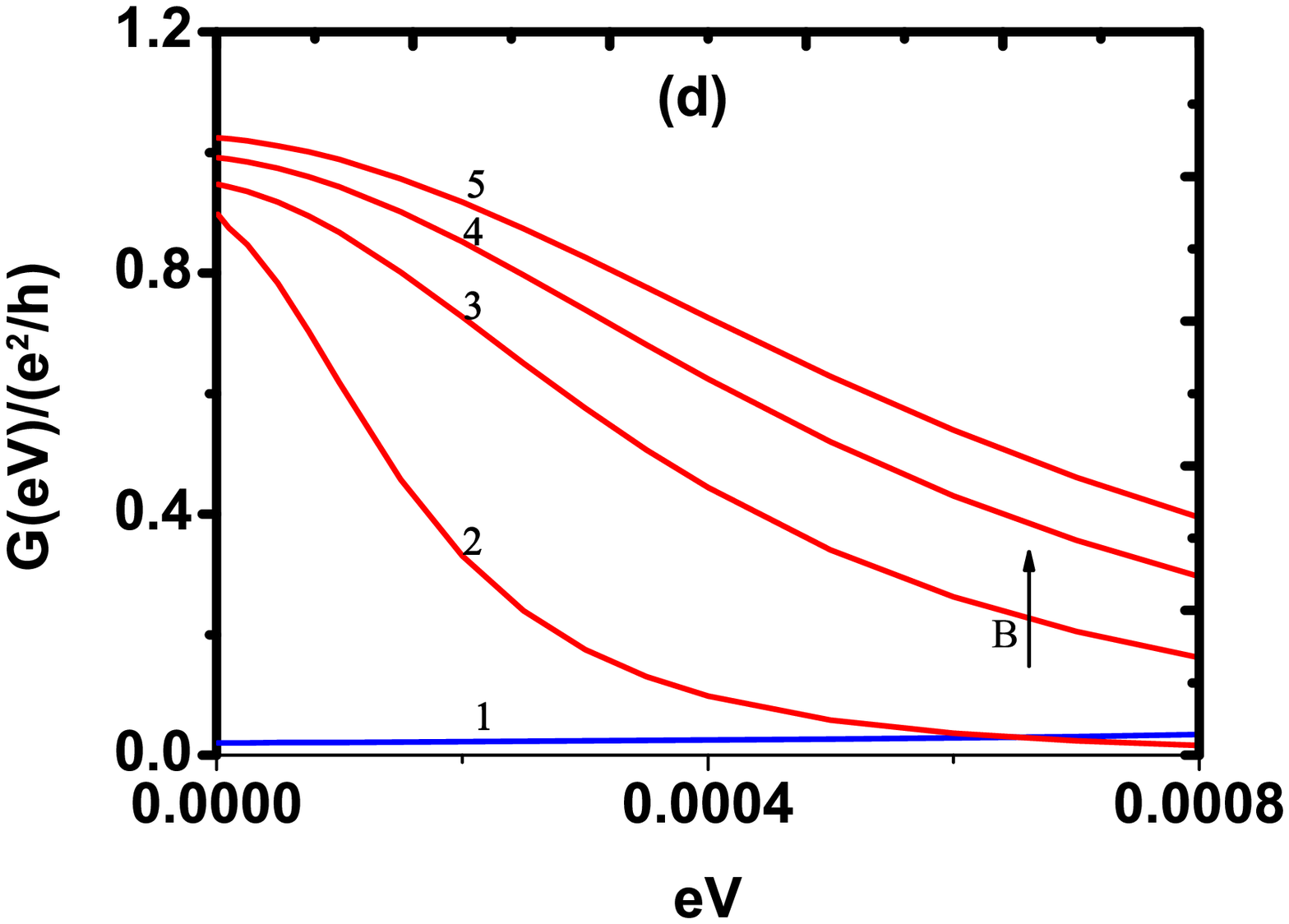}}
\caption{ (color online) Dimensionless experimental parameters:
$\mu_{n} = 1$, $\mu_{s} = 0$, $\Delta = 0.0125$, $\alpha = 0.1$,
(a) $L = \infty$, $B = 0.0135$, $\Delta_{g} = 0.001$, $Z = 1$, (b)
$L = \infty$, $B = 1$,  $\Delta_{g} = 0.001$, $Z = 1$, parameters
unshown, line $1$(black), line $3$(red) and line $5$(black): $B =
0.0135$, line 2(blue): $B = 0.006$, line 4(blue): $B = 0.026$.
Such a choice of magnetic field is to make $\Delta_{g}$ for every
line equal. (c) $B = 0.0135$, $Z = 1$, $\Delta_{g} = 0.001$. (d)
$L = 600$, $Z = 1$, $1 \rightarrow 5$, $B = 0.0115 \rightarrow
0.0195$, $\delta B = 0.002$ for two neighboring lines.}
\label{fig6}
\end{figure}

To discuss the effects of the pairing potential to the tunneling
potential, here we adopt the experimental parameters: $m = 0.015
m_{e}$, $m \alpha_{r}^{2}/2 = 50 \mu$eV, $\Delta_{r} = 0.25$meV,
$\tilde{\mu}_{s}=0$ \cite{V. Mourik}, and we choose $\mu_{n} =
20$meV. Fig.\ref{fig6}(a) shows the tunneling spectroscopy at
zero-temperature. Compared the zero-bias conductance peak with the
experiments measured value $\sim 0.1 e^{2}/h$, here the result
should still be several times larger even we consider the
temperature's smearing effect. However, as discussed before,
decreasing spin-orbit coupling (intermediate region) has great
reduction effect to the peak height. In Fig.\ref{fig6}(b), it is
shown halving the spin-orbit coupling almost corresponds to
halving the peak height. Therefore, if the spin-orbit coupling is
several times smaller than the reported one, the height of the
zero-bias conductance peak will decrease to be comparable with the
experiments measured value. Fig.\ref{fig6}(b) also shows that
decreasing the pairing potential, the peak height is greatly
increased. For sufficient small pairing potential, the zero-bias
conductance peak is found almost quantized. This result seems
counterintuitive, as the pairing potential is a necessity to
induce the topological superconductor. This confusion can be
clarified by the fact that when $\Delta < V_{z} << m
\alpha^{2}/2$, the upper band's effect is negligible, as a result,
the system is an ``effective $p$-wave" superconductor \cite{J.
Alicea1}. This suggests to observe a more striking peak in
experiments, it is better to choose a relative weaker pairing
potential proximity superconductor.

The effects of finite length $L$ for $N-hS$ junction is similar to
the $N-pS$ junction, that is, a conductance peak will locate at
finite-bias voltage when $L$ is not sufficiently long, and with
the increase of $L$, the peak moves toward to the zero-bias
voltage, see Fig.\ref{fig6}(c). For sufficiently long wire, we
find that, increasing the magnetic field, the width of the
zero-bias conductance peak will be greatly widened, see
Fig.\ref{fig6}(d).

\section{Discussion and conclusion}
\label{sec3}

In this work, based on the BTK method, we give a thorough study of
the tunneling spectroscopy of $N-pS$ junction and $N-hS$ junction.
Comparing the tunneling spectroscopy of $N-pS$ junction with
$N-hS$ junction's, we find zero-bias conductance peak appears in
both systems when their topological criterions are satisfied,
respectively. However, contrary to the stable quantized zero-bias
peak of the $N-pS$ junction, the zero-bias conductance peak of the
$N-hS$ junction is non-quantized and sensitive to parameters. The
non-quantization of the zero-bias conductance peak does not mean
the Majorana end state is absent, its existence is guaranteed by
the nontrivial topology of the bulk and the bulk-edge
correspondence \cite{Y. Hatsugai}. The non-quantization only
indicates that there are more transport channels compared to the
$N-pS$ junction. For small and intermediate spin-orbit coupling,
we find the additional channels have important contributions to
the transport. This indicates the perfect equal-spin Andreev
reflection is absent, and the heterostructure superconductor
always has a considerable gap to a truly $p$-wave superconductor.
And therefore, even the topological criterion is satisfied, using
a Majorana chain to denote the heterostructure superconductor and
then based on the ``interface electron-Majorana hopping" model
\cite{K. T. Law, Karsten Flensberg}, expecting a quantized
zero-bias conductance peak to emerge, is usually unjustified.

For strong spin-orbit coupling, we find that although, for weak
interface scattering potential, the zero-bias conductance peak is
very small and monotonically decreasing with increasing spin-orbit
coupling, a very strong interface scattering potential can
suppress the additional transport channels effectively and make
the equal-spin Andreev reflection get very close to the perfect
level, with the zero-bias conductance peak approaching the
quantized value. However, a strong spin-orbit coupling is hard to
realize, and a very strong interface scattering potential also
makes the width of the zero-bias conductance peak very narrow,
which will make detection difficult. Contrary to a combination of
strong spin-orbit coupling and strong interface scattering
potential, decreasing the pairing potential is a much more
effective way to suppress the additional channels and enhance the
zero-bias conductance peak to the quantized value.

Besides a zero-bias peak appearing when the system is driven into
the topological phase, there are another two common features
appearing in the figures. The first one is when the system goes
from the normal phase to the topological phase, the equal-spin
Andreev reflection amplitude at the neighborhood of zero-bias
voltage always has a sudden and comparatively large increment.
This indicates that in the topological region, the equal-spin
pairing becomes important, and the zero-bias conductance peak must
be related to the equal-spin pairing at zero-bias voltage. From
Fig.\ref{3}, we also see, the larger $A_{\uparrow}(eV=0)$ is, the
higher the peak is. A quantized zero-bias conductance peak always
corresponds to a perfect equal-spin Andreev reflection, $i.e.$,
$A_{\uparrow}(eV=0) = 1$. This correspondence indicates that
$A_{\uparrow}(eV=0)$ can be used as a measure of the ``effective".
A larger $A_{\uparrow}(eV=0)$ implies the equal-spin pairing
becomes more favored, and this indicates the heterostructure turns
to be a more ``effective $p$-wave superconductor". The second one
is when the magnetic field is turned on, the discontinuity of the
tunneling spectroscopy at induced-gap boundary is greatly
softened. Such a ``soft effect" will make the position of the
induced-gap boundary hard to detect, and as a result, the gap
closure is also hard to detect. Therefore, the ``soft effect"
induced by magnetic field can be supplied as a possible
explanation of the missing observation of the gap closure. Soft
gap induced by other reasons was discussed in detail in
Ref.\cite{So Takei}.

In conclusion, adopting the experimental parameters, we compare
the tunneling spectroscopy obtained with the experiment and find
that, even without consideration of effects due to disorder,
subbands and other inhomogeneities, the observation of a
non-quantized value under the experimental parameters is a natural
result. And furthermore, a spin-orbit coupling several times
smaller than the reported one in experiment, which can be taken as
a possible explanation of the quite small zero-bias conductance
observed in experiments. We suggest that, to observe a more
striking zero-bias conductance peak in future experiments, a
weaker pairing potential proximity superconductor is probably a
better choice.

\section*{Acknowledgments}
This work is supported by NSFC. Grant No.11275180.

\section*{Appendix: Quantized zero-bias conductance peak of the $N-pS$ junction}

For $N-pS$ junction, the velocity operator $v_{s}$ and $v_{n}$ are
given as ($\hbar=m=1$),

\begin{eqnarray}
v_{s}&=\frac{\partial H_{S}}{\partial k}=\left(\begin{array}{cc}
                                  k & \Delta \\
                                  \Delta & -k
                                \end{array}\right)=-i\left(\begin{array}{cc}
                                  \partial_{x} & i\Delta \\
                                  i\Delta & -\partial_{x}
                                \end{array}\right),\nonumber\\
v_{n}&=\frac{\partial H_{N}}{\partial k}=\left(\begin{array}{cc}
                                  k & 0 \\
                                 0 & -k
                                \end{array}\right)=-i\left(\begin{array}{cc}
                                  \partial_{x} & 0 \\
                                  0 & -\partial_{x}
                                \end{array}\right).
\end{eqnarray}
For $E=0$, the wave function in the $p$-wave superconductor (here
we consider the length is infinity) takes a simpler form,
\begin{eqnarray}
\psi_{S}(x)&=c\left(\begin{array}{c}
        i \\
      1
      \end{array}\right)e^{-k_{+}x}
       +d\left(\begin{array}{c}
        i \\
       1
      \end{array}\right)e^{-k_{-}x},
\end{eqnarray}
where $k_{\pm} = \sqrt{2 (\Delta^{2} - \mu_{s}) \pm 2
\sqrt{\Delta^{2}(\Delta^{2}-2\mu_{s})}}$. The wave function in the
normal lead takes the form
\begin{eqnarray}
\psi_{N}(x)=\left( \begin{array}{c}
        e^{2iqx} + b \\
      0
      \end{array} \right) e^{-iqx} + a \left( \begin{array}{c}
        0 \\
      1
      \end{array} \right) e^{iqx},
\end{eqnarray}
here $q_{e}=q_{h}=\sqrt{2\mu_{n}}$, and we use $q$ to denote both
of them. By matching the two wave function according to the
boundary conditions (\ref{6}), we obtain
\begin{eqnarray}
&&1+b=i(c+d), \nonumber \\
&&a=(c+d), \nonumber\\
&&(\Delta-k_{+})c+(\Delta-k_{-})d-q(1-b)=-iZ(1+b), \nonumber\\
&&i(\Delta-k_{+})c+i(\Delta-k_{-})d+qa=iZa.
\end{eqnarray}
A direct calculation gives
\begin{eqnarray}
&&a=-i, \nonumber\\
&&b=0,\nonumber\\
&&c=-\frac{q+i(\Delta-k_{-}-Z)}{k_{+}-k_{-}}, \nonumber\\
&&d=\frac{q+i(\Delta-k_{+}-Z)}{k_{+}-k_{-}},
\end{eqnarray}
$b=0$ indicates no normal reflection and $a=-i$ indicates a
perfect equal-spin Andreev reflection. According to the formula
(\ref{7}), the differential tunneling conductance at $E=0$ is
\begin{eqnarray}
G(0)&=&\frac{e^{2}}{h} \left[1 + A(0) - B(0) \right] \nonumber\\
&=& \frac{e^{2}}{h} \left[1 + |a|^{2} - |b|^{2} \right] \nonumber \\
&=& 2 \frac{e^{2}}{h},
\end{eqnarray}
the zero-bias conductance is quantized and independent of the interface
scattering potential.

\end{document}